\documentclass[12pt,a4paper]{article}
\usepackage{epsfig}
\usepackage{mathtools}
\usepackage{amsfonts}
\usepackage{amssymb}
\usepackage{amsmath}
\usepackage{color}
\usepackage{slashed}
\usepackage{cite}
\usepackage{caption}
\usepackage{subcaption}

\usepackage{tikz}
\usetikzlibrary{shapes}
\usetikzlibrary{trees}
\usetikzlibrary{calc}
\usetikzlibrary{decorations.pathmorphing}
\usetikzlibrary{decorations.markings}

\def\la              {\langle}
\def\ra              {\rangle}

\newcommand{\dotalpha}{{\dot{\alpha}}}

\newcommand{\abr}[1]{\langle #1 \rangle}

\newcommand{\vll}{{\smash{\lambda}}}
\newcommand{\vlt}{{\smash{\tilde{\lambda}}}}
\newcommand{\vlet}{{\smash{\tilde{\eta}}}}

\newcommand{\vllu}{\smash{\underline{\smash{\lambda}}}}
\newcommand{\vltu}{\smash{\underline{\smash{\tilde{\lambda}}}}}
\newcommand{\vleu}{\smash{\underline{\smash{\tilde{\eta}}}} }

\newcommand{\vlluu}{\smash{\underline{\underline{\smash{\lambda}}}}}
\newcommand{\vltuu}{\smash{\underline{\underline{\smash{\tilde{\lambda}}}}}}
\newcommand{\vleuu}{\smash{\underline{\underline{\smash{\tilde{\eta}}}}}}

\newcommand{\eqndot}{\, . }
\newcommand{\eqncom}{\, , }

\definecolor{grayn}{gray}{0.7}
\definecolor{lightgrayn}{gray}{0.8}

\def\bridgedistance{0.75}
\def\vacuumheight{1}
\def\ddist{0.75}
\def\hdist{\ddist*0.70710678118}
\def\labelvdist{0.3}
\def\labelhdist{0.3}
\def\labelddist{\labelvdist*0.70710678118}

\newlength{\offshellradius}
\newlength{\vacuumradius}
\newlength{\onshellradius}
\tikzstyle{db}=[circle, black, fill=black, minimum width=\onshellradius, draw, inner sep=0pt]
\tikzstyle{dw}=[circle, black, fill=white, minimum width=\onshellradius, draw, inner sep=0pt]
\tikzstyle{dvac}=[circle, black, fill=lightgrayn, minimum width=\vacuumradius, inner sep=0pt]
\tikzstyle{offshell}=[circle, black, fill=black, minimum width=\offshellradius, inner sep=0pt]
\tikzstyle{dl}=[circle, black, fill=white, inner sep=2pt]

\tikzset{
	gluon/.style={decorate, decoration={coil, amplitude=2pt, segment length=3.5pt, aspect=1}, draw=black}
}

\newcommand{\drawminimalA}[1]{
	\draw[very thick] (#1-0.5,-0) -- (#1-0.5,-0.5); % Minimal form factor
	\draw (#1-0.5,-0.5) -- (#1-1,-\vacuumheight);
	\draw (#1-0.5,-0.5) -- (#1,-\vacuumheight);}

\newcommand{\drawvacp}[1]{
	\draw (#1-1,-.25) -- (#1-1,-\vacuumheight); % Vacuum
	\node[dvac] at (#1-1,-0.25) {$+$};}
\newcommand{\drawvacm}[1]{
	\draw (#1-1,-.25) -- (#1-1,-\vacuumheight); % Vacuum
	\node[dvac] at (#1-1,-0.25) {$-$};}
\newcommand{\drawbridge}[2]{
	\draw (#1-1,-\vacuumheight -#2*\bridgedistance+\bridgedistance) -- (#1,-\vacuumheight -#2*\bridgedistance+\bridgedistance); % First bridge
	\node[dw] at (#1-1,-\vacuumheight -#2*\bridgedistance+\bridgedistance) {};
	\node[db] at (#1,-\vacuumheight -#2*\bridgedistance+\bridgedistance) {};}
\newcommand{\drawvline}[2]{
	\draw (#1-1,-\vacuumheight) -- (#1-1,-\vacuumheight -#2*\bridgedistance);}
\newcommand{\drawnumline}[2]{
	\draw (#1-1,-.5) -- (#1-1,-1.5*\vacuumheight); % Vacuum
	\node at (#1-1,-0.25) {#2};}
\newcommand{\drawoffshellbridge}[2]{
	\draw (#1-1,-\vacuumheight -#2*\bridgedistance+\bridgedistance) -- (#1,-\vacuumheight -#2*\bridgedistance+\bridgedistance); % First bridge
	\node[offshell] at (#1-1,-\vacuumheight -#2*\bridgedistance+\bridgedistance) {};
	\node[offshell] at (#1,-\vacuumheight -#2*\bridgedistance+\bridgedistance) {};}

\begin{document}

\begin{center}

\vspace{1cm}

{\bf \Large Grassmannian integral for general gauge invariant off-shell amplitudes in $\mathcal{N}=4$ SYM.} \vspace{1cm}

{\large L.V. Bork$^{1,2}$ A.I. Onishchenko$^{3,4,5}$}\vspace{0.5cm}

{\it $^1$Institute for Theoretical and Experimental Physics, Moscow,
	Russia,\\
	$^2$The Center for Fundamental and Applied Research, All-Russia
	Research Institute of Automatics, Moscow, Russia, \\
	$^3$Bogoliubov Laboratory of Theoretical Physics, Joint
	Institute for Nuclear Research, Dubna, Russia, \\
	$^4$Moscow Institute of Physics and Technology (State University), Dolgoprudny, Russia, \\
	$^5$Skobeltsyn Institute of Nuclear Physics, Moscow State University, Moscow, Russia}\vspace{1cm}

\abstract{In this paper we consider tree-level gauge invariant off-shell amplitudes (Wilson line form factors) in $\mathcal{N}=4$ SYM with arbitrary number of off-shell gluons or equivalently Wilson line operator insertions. We make a conjecture for the Grassmannian integral representation for such
objects and verify our conjecture on several examples. It is remarkable that in our formulation
one can consider situation when on-shell particles are not present at all, i.e. we have Grassmannian integral representation for purely off-shell object.
In addition we show that off-shell amplitude with arbitrary number of off-shell gluons could be also obtained using quantum inverse scattering method for auxiliary $\mathfrak{gl}(4|4)$ super spin chain.}
\end{center}

\begin{center}
\end{center}

\newpage

\tableofcontents{}\vspace{0.5cm}

\renewcommand{\theequation}{\thesection.\arabic{equation}}

\section{Introduction}

Following Witten's twistor string theory \cite{WittenTwistorStrings} we have witnessed an enormous progress in understanding the structure of S-matrix of $\mathcal{N}=4$ SYM  as well as other gauge theories (see \cite{Reviews_Ampl_General,Henrietta_Amplitudes} for a review). The central role in these achievements was played by newly developed computational methods, such as BCFW recursion \cite{BCFW1,BCFW2} for tree amplitudes and generalized unitarity (see \cite{Reviews_Ampl_General} and references therein) for loop amplitudes. The power of the mentioned \emph{on-shell methods} is largely due to the introduction of new variables, such as helicity spinors, momentum twistors \cite{MomentumTwistors}, link variables \cite{linkvariables} together with an extensive use of superspace methods \cite{Nair,DualSuperConformalSymmetry}. The use of on-shell methods resulted in the explicit answers for amplitudes both at high orders of perturbation theory and/or with large number of external legs (see  \cite{Reviews_Ampl_General,Henrietta_Amplitudes} for a review and reference therein). The latter have led to several important all-loop conjectures as well as to the discovery of underlying integrable structure of $\mathcal{N}=4$ SYM S-matrix
\cite{YangianSymmetryTreeAmplitudes,BeisertYangianRev,Beisert_SpectralReg_New,AmplitudesSpectralParameter1,AmplitudesSpectralParameter2,Chicherin_YangBaxterScatteringAmplitudes,Frassek_BetheAnsatzYangianInvariants,Kanning_ShortcutAmplitudesIntegrability,Broedel_DictionaryRoperatorsOnshellGraphsYangianAlgebras,Broedel_DeformedOneLoopAmplitudes}. We should mention an important research direction originated with the connected RSV prescription \cite{RSV,RSVlinkvariables}. Within the latter  $\mathcal{N}=4$ SYM amplitudes are expressed in terms of the localized integrals over the moduli space of $n$-punched Riemann spheres. The next progress along this direction was due to the introduction of scattering equations \cite{scattering-eq1,scattering-eq2,scattering-eq3,scattering-eq4,scattering-eq5}, which were later generalized to loop level \cite{ambitwistor1,ambitwistor2,ambitwistor3,ambitwistor4} and derived from ambitwistor string theory \cite{ambitwistor5}.

Another novel and important direction in the study of scattering amplitudes is related to their Grassmannian integral representation
\cite{DualitySMatrix,AmplitudesPositiveGrassmannian,AllLoopIntegrandN4SYM,GrassmanianOriginDualConformalInvariance,UnificationResiduesGrassmannianDualities,DualSupercondormalInvarianceMomentumTwistorsGrassmannians}. It does not only naturally unifies different BCFW representations for tree level amplitudes and loop level integrands  \cite{DualitySMatrix,AmplitudesPositiveGrassmannian}, but also played a key role in the discovery and study of the integrable structure behind $\mathcal{N}=4$ SYM  S-matrix  \cite{Drummond_Grassmannians_Tduality,Drummond_Yangian_origin_Grassmannian_integral,Chicherin_YangBaxterScatteringAmplitudes}. Moreover, this Grassmannian integral representation also naturally relates perturbative $\mathcal{N}=4$ SYM and twistor string theories amplitudes\cite{UnificationResiduesGrassmannianDualities}.
In addition  a possible geometrical interpretation of $\mathcal{N}=4$ SYM S-matrix (so called Amplituhedron) was discovered just with the use of Grassmannian picture
\cite{MomentumTwistors,Arcani_Hamed_Polytopes,Amplituhdron_1,Amplituhdron_2,Amplituhdron_3,Amplituhdron_4,Amplituhdron_5,Amplituhdron_6}.

The on-shell methods were also successfully applied to partially off-shell objects, such as form factors in $\mathcal{N}=4$ SYM theory  \cite{FormFactorMHV_component_Brandhuber,HarmonyofFF_Brandhuber,BORK_NMHV_FF,FF_MHV_3_2loop,Roiban_FormFactorsOfMultipleOperators,FormFactorMHV_half_BPS_Brandhuber,FormFactorMHV_Remainder_half_BPS_Brandhuber,BORK_POLY,Wilhelm_Integrability_1,Wilhelm_Integrability_2,Wilhelm_Integrability_3,Wilhelm_Integrability_4,Wilhelm_Twisors_1,Wilhelm_Twisors_2,LHC_1,LHC_2,LHC_3,BoFeng_BoundaryContributions,Henn_Different_Reg_FF,3loopSudakovN4SYM,FF_Colour_Kinematic,masters4loopSudakovN4SYM,Oluf_Tang_Engelund_Lagrangian_Insertion,Brandhuber_DynamicFormFactors,BrandhuberConnectedPrescription,HeConnectedPrescription,Sudakov5loop,PenanteThesis,FormFactorsMaximumTrascedentality}.
The latter may be viewed as the amplitudes of the processes, in which classical field coupled through gauge invariant operator $\mathcal{O}$ produces an on-shell quantum state. Grassmannian representation is no exception and can be applied to form factors as well \cite{SoftTheoremsFormFactors,FormFactorsGrassmanians,WilhelmThesis,q2zeroFormFactors}.

Another interesting off-shell objects, which will be the subject of the present paper, are gauge invariant off-shell amplitudes  \cite{vanHamerenBCFW1,vanHamerenBCFW2,LipatovEL1,LipatovEL2,KirschnerEL1,KirschnerEL2,KotkoWilsonLines,vanHamerenWL1,vanHamerenWL2} (also known as reggeon amplitudes within the context of Lipatov's effective lagrangian), which one encounters within
$k_T$ - or high-energy factorization approach
\cite{GribovLevinRyskin,CataniCiafaloniHautmann,CollinsEllis,CataniHautmann} as well as in the study of processes at multi-regge kinematics. For other studies of off-shell currents and amplitudes see \cite{BerendsGiele,recursion_solution,offsellBCFWcurrents,Kallosh_1,Kallosh_2}. In the study of form factors we are dealing with local
gauge invariant color singlet operators, for example operators from stress-tensor operator supermultiplet  \cite{FormFactorMHV_component_Brandhuber,HarmonyofFF_Brandhuber,BORK_NMHV_FF,FF_MHV_3_2loop,Roiban_FormFactorsOfMultipleOperators,BKV_Form_Factors_N=4SYM,BKV_SuperForm,Zhiboedov_Strong_coupling_FF,Strong_coupling_FF_Yang_Gao}. However, we may consider also gauge invariant non-local operators, for example  Wilson loops (lines) or their products.  This way we may consider \emph{gauge invariant off-shell amplitudes} as form factors of
Wilson line operators or their products  \cite{vanHamerenBCFW1,vanHamerenBCFW2,LipatovEL1,LipatovEL2,KirschnerEL1,KirschnerEL2,KotkoWilsonLines,vanHamerenWL1,vanHamerenWL2}.
An insertion of Wilson line operator corresponds to the off-shell or reggeized gluon in such formulation. In our previous paper \cite{offshell-1leg} we presented a conjecture for Grassmannian integral representation of gauge invariant off-shell amplitudes with one leg off-shell and shown how they could be described in a language of auxiliary  $\mathfrak{gl}(4|4)$ super spin chain. The purpose of this paper is to extend the results of \cite{offshell-1leg} to the case of amplitudes with multiple off-shell gluons.

This paper is organized as follows. In section 2 after recalling necessary definitions we formulate a conjecture for Grassmannian integral representation of gauge invariant amplitudes with multiple off-shell gluons. Here we also formulate a hypothesis for the structure of on-shell diagrams for off-shell amplitudes. The appendix \ref{appA} contains a check of the latter in the case of 3-point amplitude with two off-shell gluons. In section 3 we use explicit examples with known BCFW answers \cite{vanHamerenBCFW1} to perform checks of our conjecture. The explicit calculations are performed using both spinor helicity and momentum twistor representations. Section 4 is devoted to the auxiliary spin chain description of the off-shell amplitudes and finally we come with our conclusion.

\section{Gauge invariant off-shell amplitudes and regulated integral over Grassmannian}

One way to define gauge invariant amplitudes containing off-shell gluons is in terms of form factors of Wilson line operators \cite{KotkoWilsonLines}:
\begin{eqnarray}\label{WilsonLineOperDef}
\mathcal{W}_p^c(k) = \int d^4 x e^{ix\cdot k} \mathrm{Tr} \left\{
\frac{1}{\pi g} t^c \; \mathcal{P} \exp\left[\frac{ig}{\sqrt{2}}\int_{-\infty}^{\infty}
ds \; p\cdot A_b (x+ sp) t^b\right]
\right\}.
\end{eqnarray}
Here $t^c$ are $SU(N_c)$ generators\footnote{The color generators are normalized as $\mathrm{Tr} (t^a t^b) = \delta^{a b}$} and we also assumed so called  $k_T$ - decomposition of the off-shell gluon momentum $k$, $k^2 \neq 0$:
\begin{eqnarray}\label{kT}
k^{\mu} = x p^{\mu} + k_T^{\mu},
\end{eqnarray}
where $p$ is the gluon direction (also known as the off-shell gluon polarization vector), such that $p^2=0$, $p\cdot k = 0$ and $x \in [0,1]$. There is a freedom in such decomposition, which could be parametrized by an auxiliary light-like four-vector $q^{\mu}$, so that
\begin{eqnarray}
k_T^{\mu} (q) = k^{\mu} - x(q) p^{\mu}\quad \text{with}\quad x(q) = \frac{q\cdot k}{q\cdot p} \;\; \text{and} \;\; q^2 = 0.
\end{eqnarray}
Using the fact, that now $k_T^{\mu}$ is transverse both with respect to $p^\mu$ and $q^\mu$ vectors, the off-shell gluon transverse momentum $k_T^{\mu}$ could be expanded in the basis of two ``polarization'' vectors\footnote{Here we used helicity spinor decomposition of light-like four-vectors $p$ and $q$.}  as \cite{vanHamerenBCFW1}:
\begin{eqnarray}
k_T^{\mu} (q) = -\frac{\kappa}{2}\frac{\la p|\gamma^{\mu}|q]}{[pq]}
- \frac{\kappa^{*}}{2}\frac{\la q|\gamma^{\mu}|p]}{\la qp\ra}\quad
\text{with} \quad \kappa = \frac{\la q|\slashed{k}|p]}{\la qp\ra},\;
\kappa^{*} = \frac{\la p|\slashed{k}|q]}{[pq]}.
\end{eqnarray}
It is easy to see, that $k^2 = -\kappa\kappa^{*}$ and using Schouten identities it could be shown, that both $\kappa$ and $\kappa^{*}$ are independent of auxiliary four-vector $q^{\mu}$ \cite{vanHamerenBCFW1}. Another useful relation, which follows directly from $k_T$ decomposition and will be used often in what follows, is given by
\begin{eqnarray}\label{pBracetsRelation}
	k|p\rangle=|p]\kappa^*.
\end{eqnarray}
Note, that Wilson line operator we use to describe off-shell gluon is colored. It is invariant $\delta\mathcal{W}_p^c(k) = 0$ under local infinitesimal gauge transformations $\delta A_{\mu} = [D_{\mu},\chi]$ with $\chi$ vanishing at infinity $x\to\infty$.  At the same time it transform in the adjoint representation under global $SU (N_c)$ transformations with constant $\chi$ as \cite{LipatovEL1,LipatovEL2}:
\begin{eqnarray}
	\delta\mathcal{W}_p(k) = g [\mathcal{W}_p(k), \chi] .
\end{eqnarray}

Using Wilson line operator defined above the gauge invariant  amplitude with one off-shell and $n$ on-shell gluons can be written as \cite{KotkoWilsonLines}:
%\begin{eqnarray}\label{AmplitudeOnelOffShellGluon}
%\mathcal{A}_{n+1} \left(1^{\pm},\ldots ,n^{\pm},(n+1)^*\right) = \la k_1, \epsilon_1, c_1;\ldots
%;k_n,\epsilon_n,c_n|\mathcal{W}_p^{c_{n+1}}(k)|0\ra
%\end{eqnarray}
\begin{eqnarray}\label{AmplitudeOnelOffShellGluon}
	\mathcal{A}_{n+1} \left(1^{\pm},\ldots ,n^{\pm},g_{n+1}^*\right) = \la \{k_i, \epsilon_i, c_i\}_{i=1}^m|\mathcal{W}_p^{c_{n+1}}(k)|0\ra\,
\end{eqnarray}
Here asterisk denotes an off-shell gluon and $p$, $k$, $c$ are its direction, momentum and color index correspondingly. Next $\la \{k_i, \epsilon_i, c_i\}_{i=1}^m|=\bigotimes_{i=1}^m\la k_i,\varepsilon_i, c_i|$ and $\la k_i,\varepsilon_i, c_i|$ denotes on-shell gluon state  with momentum $k_i$, polarization vector $\varepsilon_i^-$ or $\varepsilon_i^+$ and color index $c_i$. Amplitudes with multiple off-shell gluons can be represented in a similar fashion:
%\begin{eqnarray}\label{AmplitudeSeveralOffShellGluons}
%&&\mathcal{A}_{m+n} \left(1,\ldots ,m,(m+1)^*\ldots (n+m)^*\right) =\nonumber\\
%&&\la k_1, \epsilon_1, c_1;\ldots
%;k_m,\epsilon_m,c_m|\mathcal{W}_{p_{m+1}}^{c_{m+1}}(k_{m+1})\ldots \mathcal{W}_{p_{n+m}}^{c_{n+m}}(k_{n+m})|0\ra,
%\end{eqnarray}
\begin{eqnarray}\label{AmplitudeSeveralOffShellGluons}
	\mathcal{A}_{m+n} \left(1^{\pm},\ldots ,m^{\pm},g_{m+1}^*,\ldots ,g_{n+m}^*\right) =
	\la \{k_i, \epsilon_i, c_i\}_{i=1}^m |\prod_{j=1}^n\mathcal{W}_{p_{m+j}}^{c_{m+j}}(k_{m+j})|0\ra,
\end{eqnarray}
where $p_i$ is the direction of the $i$'th ($i=1,...,n$) off-shell gluon and $k_i$ is its momentum. As a function of kinematical variables this function is given by
\begin{eqnarray}
\mathcal{A}_{m+n} \left(1^{\pm},\ldots ,g_{n+m}^*\right) =\mathcal{A}_{m+n}\left(\{\lambda_i,\tilde{\lambda}_i,\pm,c_i\}_{i=1}^{m};
\{k_j,\lambda_{p,j},\tilde{\lambda}_{p,j},c_j\}_{j=m+1}^{m+n}\right),
\end{eqnarray}
where $\lambda_{p,j},\tilde{\lambda}_{p,j}$ are spinors coming from helicity spinor decomposition of of $j$'th Wilson line direction vector $p_j$. Note, that we can consider a situation where only off-shell gluons are present
(correlation function of Wilson line operators):
\begin{eqnarray}\label{CorrFunctionOffShellGluons}
\mathcal{A}_{0+n} \left(g_1^*\ldots g_n^*\right) =
\la 0|\mathcal{W}_{p_{1}}^{c_{1}}(k_{1})\ldots \mathcal{W}_{p_{n}}^{c_{n}}(k_{n})|0\ra.
\end{eqnarray}
In practical calculations it is more convenient to deal with the color ordered versions of the above amplitudes.  The original amplitudes are then recovered through the color decomposition\footnote{See for example \cite{offshell-1leg,DixonReview}.}:
%\begin{eqnarray}\label{ColourOrderedAmplitudeDefinition}
%\mathcal{A}_n^{tree}({p_i,h_i,a_i}) = g^{n-2}\sum_{\sigma \in S_n/Z_n} \text{tr}
%~(t^{a_{\sigma (1)}}\cdots t^{a_{\sigma (n)}}) A_n^{tree} (\sigma(1^{h_1}),\ldots ,\sigma (n^{h_n})),
%\end{eqnarray}
\begin{eqnarray}\label{ColourOrderedAmplitudeDefinition}
	\mathcal{A}_{n+m}^*(1^{\pm},\ldots ,m^{\pm},g_{m+1}^*,\ldots, g_{m+n}^*) &=& g^{n-2}\sum_{\sigma \in S_{n+m}/Z_{n+m}} \text{tr}
	~(t^{a_{\sigma (1)}}\cdots t^{a_{\sigma (n)}})\times\nonumber\\
 &\times&A_{n+m}^* \left(\sigma(1^{\pm}),\ldots, \sigma(g_{n+m}^*))\right).
\end{eqnarray}
%where $g$ is the gauge coupling, $p_i$, $h_i$ are the gluon momenta and helicities.
Here $S_{n+m}$ is the set of all permutations of $n+m$ objects and $Z_{n+m}$ is the subset of cyclic permutations. We will also sometimes write $g_i^{\pm}$ instead of $i^{\pm}$ to denote on-shell gluon to make some formulas more transparent.

In the case of $\mathcal{N}=4$ SYM one can also consider other then gluons on-shell states from $\mathcal{N}=4$ supermultiplet. The way to do it is to combine all sixteen on-shell states of $\mathcal{N}=4$ SYM into one on-shell chiral superfield \cite{Nair}:
\begin{eqnarray}
\Omega= g^+ + \vlet_A\psi^A + \frac{1}{2!}\vlet_{A}\vlet_{B}\phi^{AB}
+ \frac{1}{3!}\vlet_A\vlet_B\vlet_C\epsilon^{ABCD}\bar{\psi}_{D}
+ \frac{1}{4!}\vlet_A\vlet_B\vlet_C\vlet_D\epsilon^{ABCD}g^{-},
\end{eqnarray}
where $g^+, g^-$ denote creation/annihilation operators of gluons with $+1$ and $-1$ hecilities, $\psi^A$ are creation/annihilation operators of four Weyl spinors with negative helicity $-1/2$, $\bar{\psi}_A$ are creation/annihilation operators of four Weyl spinors with positive helicity and $\phi^{AB}$ stand for creation/annihilation operators of six scalars (anti-symmetric in the $SU(4)_R$ $R$-symmetry  indices $AB$).  In what follows we will also need superstates defined by the action of superfield creation/annihilation operators on vacuum. For $n$-particle superstate we have
$\la \Omega_1\Omega_2\ldots\Omega_n|\equiv \bigotimes_{i=1}^n\la 0|\Omega_i$ and corresponding  $\mathcal{N}=4$ SYM superamplitudes could be written as
\begin{eqnarray}\label{AmplitudeSeveralOffShellGluonsSUSY}
	A_{m+n}^* \left(\Omega_1,\ldots,\Omega_m,g_{m+1}^*,\ldots ,g_{n+m}^*\right) =
	\la \Omega_1\ldots\Omega_m|
\prod_{j=1}^n\mathcal{W}_{p_{m+j}}(k_{m+j})|0\ra,
\end{eqnarray}
where the explicit dependence of $A_{m+n}^* \left(\Omega_1,\ldots,g_{m+n}^*\right)$ amplitude on kinematical variables is given by
\begin{eqnarray}\label{AmplitudeSeveralOffShellGluonsArgumentsSUSY}
A_{m+n}^* \left(\Omega_1,\ldots,g_{m+n}^*\right) =A_{m+n}^*\left(\{\lambda_i,\tilde{\lambda}_i,\tilde{\eta}_i\}_{i=1}^{m};
\{k_i,\lambda_{p,i},\tilde{\lambda}_{p,i}\}_{i=m+1}^{m+n}\right).
\end{eqnarray}
This object contains not only amplitudes with on-shell gluons, but also all amplitudes with other on-shell states from $\mathcal{N}=4$ supermultiplet. The spinors $\lambda_i,\tilde{\lambda}_i$ encode kinematics of on-shell states, while $\tilde{\eta}_i$ encodes their helicity content. Off-shell momentum $k_i$ and light-cone direction vector $p_i=\lambda_{p,i}\tilde{\lambda}_{p,i}$ encode information about Wilson line operator insertion.
So, in what follows we are considering partially supersymmetrized version of (\ref{AmplitudeSeveralOffShellGluons}) with on-shell states treated in supersymmetric manner, while Wilson line operators ("off-shell states") left unsupersymmetrized. The amplitudes with gluons, scalars, etc. ("component amplitudes") can then be extracted as coefficients in $\tilde{\eta}$ expansion of $A_{m+n}^*$ amplitude similar to the case of ordinary on-shell amplitudes and super form factors.
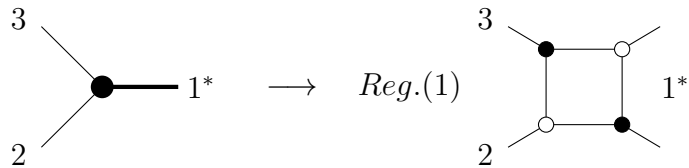
\begin{figure}[htbp]
	\begin{gather*}
	\begin{aligned}
	\begin{tikzpicture}[baseline={($(na.base) - (0,0.1)$)},transform shape,scale=1]
	\draw[ultra thick] (0,0) -- (1,0);
	\draw (0,0) -- (-0.8,0.8);
	\draw (0,0) -- (-0.8,-0.8);
	\node[offshell] (na) at (0,0) {};
	\node at (1+\labelhdist,0) {$1^*$};
	\node at (-0.8-\labelhdist,-0.9) {$2$};
	\node at (-0.8-\labelhdist,0.9) {$3$};
	\end{tikzpicture}
	\end{aligned}
	\quad\longrightarrow\quad Reg.(1)
	\begin{aligned}
	\begin{tikzpicture}[baseline={($(na.base) - (0,1)$)},transform shape, scale=1]
	\draw (0,0) -- (1,0);
	\draw (0,-1) -- (1,-1);
	\draw (0,0) -- (0,-1);
	\draw (1,0) -- (1,-1);
	\draw (0,0) -- (-0.5,0.3);
	\draw (0,-1) -- (-0.5,-1.3);
	\draw (1,0) -- (1.5,0.3);
	\draw (1,-1) -- (1.5,-1.3);
	\node[db] (na) at (0,0) {};
	\node[dw] at (1,0) {};
	\node[dw] at (0,-1) {};
	\node[db] at (1,-1) {};
	\node at (1.4 + \labelhdist,-0.5) {$1^*$};	
	\node at (-0.5 - \labelhdist,-1.4) {$2$};
	\node at (-0.5 - \labelhdist,0.4) {$3$};
	\end{tikzpicture}
	\end{aligned}
	\end{gather*}
	\caption{Off-shell 3-point vertex $A^{*}_{2+1}$
	}
	\label{fig: offshell-vertex}
\end{figure}

The color ordered versions of all types of off-shell gauge invariant amplitudes considered before could be efficiently computed using an off-shell generalization \cite{vanHamerenBCFW1,vanHamerenBCFW2} of the original on-shell BCFW recursion \cite{BCFW1,BCFW2}. Later in \cite{offshell-1leg} we presented a conjecture for Grassmannian integral representation of tree level color ordered version of (\ref{AmplitudeOnelOffShellGluon}) amplitude  $A_{k,n+1}^*$  (here\footnote{We hope that $k$ here will not be confused with the off-shell gluon momentum introduced before and its precise meaning will be always clear from the context.} $k$ is related to overall helicity $\lambda_{\Sigma}$ of on-shell particles  given by $\lambda_{\Sigma}=n+2-2k$). In our consideration in \cite{offshell-1leg} $n$ on-shell particles were treated in manifestly supersymmetric way, while the off-shell gluon remained unsupersymmetrized. To be more explicit\footnote{We refer the interested reader to \cite{offshell-1leg} for more details.}, let us consider the following
Grassmannian integral over $Gr(n+2,k)$:
\begin{equation}\label{GrassmannianIntegralForOneOffShellGluon}
\Omega_{n+2}^k[\Gamma]=
\int_{\Gamma}\frac{d^{k\times (n+2)}C}{\text{Vol}[GL (k)]}Reg.
\frac{\delta^{k\times 2} \left( C
\cdot \vltuu \right)
\delta^{k\times 4} \left(C \cdot \vleuu \right)
\delta^{(n+2-k)\times 2} \left(C^{\perp} \cdot \vlluu \right)}{(1 \cdots k)\cdots (n+1\;n+2 \cdots k-2) (n+2 \; 1\cdots k-1)},
\end{equation}
where
\begin{equation}\label{RegFunction}
Reg.=\frac{\la\xi p\ra}{\kappa^{*}}\frac{(n+2 \; 1\cdots k-1)}{(n+1 \; 1\cdots k-1)},
\end{equation}
and external kinematical variables are defined as
\begin{align}\label{SpinorsInDeformadGrassmannian}
&\vlluu_i = \vll_i, & i = 1,\ldots  n& , &\vlluu_{n+1} &= \lambda_p ,
&\vlluu_{n+2} &= \xi, \nonumber \\
&\vltuu_i = \vlt_i, & i = 1,\ldots  n& ,
&\vltuu_{n+1} &= \frac{\la\xi |k}{\la\xi p\ra},
&\vltuu_{n+2} &= - \frac{\la p |k}{\la\xi p\ra} , \nonumber \\
&\vleuu_i = \vlet_i,  & i = 1,\ldots  n& , &\vleuu_{n+1} &= \vlet_p ,
&\vleuu_{n+2} &= 0 . \nonumber \\
\end{align}
Here $k$ is the off-shell gluon momentum. It is also assumed, that the off-shell momentum $k$ is $k_T$ decomposed (\ref{kT}), so that $p=\lambda_p\tilde{\lambda}_p$ and $q=\xi\tilde{\xi}$. We claim, that making an appropriate choice of integration contour $\Gamma$,
which is likely to be given by "tree contour" $\Gamma_{tree}$ \cite{DualitySMatrix} of $n+2$ point $\mbox{N}^{k-2}\mbox{MHV}$ on-shell amplitude, the gauge invariant amplitude with one gluon off-shell and $n$ on-shell particles in $\mathcal{N}=4$ SYM could be written as \cite{offshell-1leg}:
\begin{equation}
A^*_{k,n+1}(\Omega_1,\ldots,\Omega_n,g^*_{n+1})=\frac{\partial^4}{\partial \tilde{\eta}_p^4}\Omega_{n+2}^k[\Gamma_{tree}]
\end{equation}
This relation was successfully verified for arbitrary $n$ and $k=2,3$. The factor $Reg.$ in (\ref{GrassmannianIntegralForOneOffShellGluon}) can be considered as a deformation or IR regulator of the Grassmannian integral for on-shell amplitudes. Namely, the Grassmannian integral representation of on-shell amplitude $A^k_n$ is given by ($\Gamma=\Gamma_{tree}$):
\begin{equation}
L^k_n[\Gamma]=\int_{\Gamma}\frac{d^{k\times n}C}{\text{Vol}[GL (k)]}
\frac{\delta^{k\times 2} \left(C
\cdot \tilde{\lambda}\right)
\delta^{k\times 4} \left(C \cdot \tilde{\eta} \right)
\delta^{(n-k)\times 2} \left(C^{\perp} \cdot \lambda \right)}{(1 \cdots k)\cdots
(n-1\; n \cdots k-2)(n \; 1\cdots k-1)}. \label{GrassmannianOnshell}
\end{equation}
As it is known, this integral is singular in the holomorphic soft limit (we take limit $p \rightarrow 0$ such that for $p=\lambda\tilde{\lambda}$, $\lambda \mapsto \epsilon \lambda$ and
$\tilde{\lambda} \mapsto \tilde{\lambda}$ and  $\epsilon \rightarrow 0$) with respect to any momentum of external on-shell particle $p_i$. The behavior of the integral in such limit is controlled by soft theorems \cite{Weinberg_SoftGravitonTheorem,Stromingeк_BMS_GravitationalScattering,He_BMS_SoftTheorem,Cachazo_NewSoftGravitonTheorem,Strominger_DisplacementMemory,Pasterski_SpinMemory,Strominger_AsymptoticSymmetriesYM,He_KacMoodyYM,Dumitrescu_InfiniteFermionicSymmetry,Geyer_AmbitwistorSoftTheorems,Lipstein_SoftTheoremsCFT,Laenen_PathIntegralEikonal,Casali_SoftSubLeadingYM,Bern_Soft_YM_Grav_loops,Cachazo_Soft_YM_Grav_loops,Brandhuber_OneLoopSoftTheoremsDualSuperConformSym,N4SoftTheoremsGrassmannian}. The same behavior  also holds for the holomorphic soft limit with respect to the on-shell momenta of the $\Omega_{n+2}^k$ Grassmannian integral.
The soft behavior of the amplitude with one gluon off-shell $A^*_{k,n+1}$ with respect to the holomorphic soft limits of the  on-shell variables $p$ and $q$ parameterizing off-shell gluon is however different. In this limit it must be regular and the following relation should hold\footnote{See \cite{offshell-1leg} for more details.} ($q=\xi\tilde{\xi}$,
$\epsilon  \rightarrow 0$)
\begin{equation}
A^*_{k,n+1}\Big{|}_{\xi\mapsto \epsilon \xi}=\frac{1}{\kappa^{*}}A^k_{n+1}+O(\epsilon),
\end{equation}
with the helicity of the on-shell gluon with momentum $p_{n+1}=p$ equal to $-1$. The behavior of the $\Omega_{n+2}^k$ with respect to such limit is similar \cite{offshell-1leg}:
\begin{eqnarray}
\Omega^{k}_{n+2}[\Gamma_{tree}]\Big{|}_{\xi\mapsto \epsilon \xi}=\frac{1}{\kappa^{*}}L^k_{n+1}[\Gamma_{tree}']+O(\epsilon),
\end{eqnarray}
where contours $\Gamma_{tree}$ and $\Gamma_{tree}'$ are identical except that $\Gamma_{tree}$ include poles at the zeros of the minors\footnote{See \cite{offshell-1leg} and \cite{N4SoftTheoremsGrassmannian} for details} $(n-k+4  \cdots 1)$ up to $(n+1 \cdots k-3)$. The insertion of $Reg.$ function is exactly what gives the desired behavior of $\Omega^{k}_{n+2}$ and in fact the form of $Reg.$ function can be fixed by this requirement.
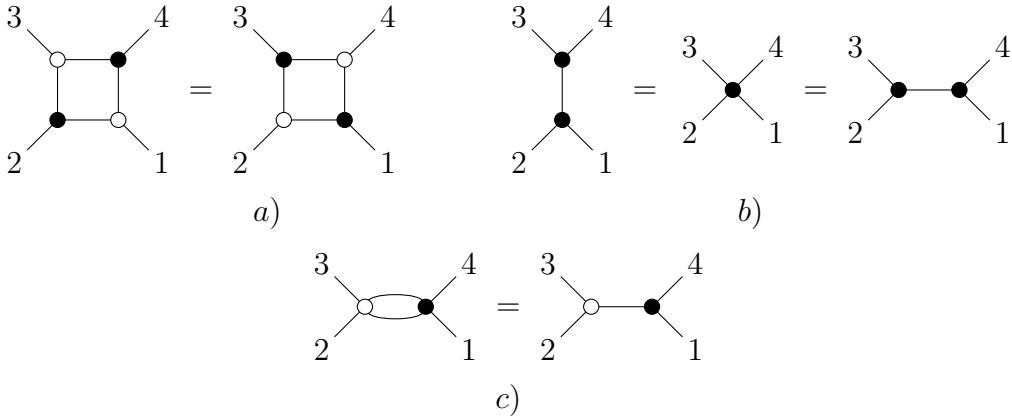
\begin{figure}[h]
	\centering
	\begin{gather*}
	\begin{tikzpicture}[baseline={($(n1.base) + (0,1)$)},scale=0.8,rotate=0]
	\draw (1,1) -- (1,2) -- (2,2) -- (2,1) -- (1,1);
	\draw (0.5,0.5) -- (1,1);
	\draw (2.5,0.5) -- (2,1);
	\draw (2.5,2.5) -- (2,2);
	\draw (0.5,2.5) -- (1,2);
	\node[dw] at (2,1) {};
	\node[dw] at (1,2) {};
	\node[db] at (1,1) {};
	\node[db] at (2,2) {};
	\node (n1) at (2.5+\labelddist,0.5-\labelddist) {1};
	\node at (0.5-\labelddist,2.5+\labelddist) {3};
	\node at (0.5-\labelddist,0.5-\labelddist) {2};
	\node at (2.5+\labelddist,2.5+\labelddist) {4};
	\end{tikzpicture}
	=
	\begin{tikzpicture}[baseline={($(n1.base) + (0,1)$)},scale=0.8,rotate=0]
	\draw (1,1) -- (1,2) -- (2,2) -- (2,1) -- (1,1);
	\draw (0.5,0.5) -- (1,1);
	\draw (2.5,0.5) -- (2,1);
	\draw (2.5,2.5) -- (2,2);
	\draw (0.5,2.5) -- (1,2);
	\node[db] at (2,1) {};
	\node[db] at (1,2) {};
	\node[dw] at (1,1) {};
	\node[dw] at (2,2) {};
	\node (n1) at (2.5+\labelddist,0.5-\labelddist) {1};
	\node at (0.5-\labelddist,2.5+\labelddist) {3};
	\node at (0.5-\labelddist,0.5-\labelddist) {2};
	\node at (2.5+\labelddist,2.5+\labelddist) {4};
	\end{tikzpicture} \qquad\quad
	\begin{tikzpicture}[baseline={($(n1.base) + (0,1)$)},scale=0.8,rotate=0]
	\draw (1,1) -- (1,2);
	\draw (0.5,0.5) -- (1,1);
	\draw (1.5,0.5) -- (1,1);
	\draw (1.5,2.5) -- (1,2);
	\draw (0.5,2.5) -- (1,2);
	\node[db] at (1,1) {};
	\node[db] at (1,2) {};
	\node (n1) at (1.5+\labelddist,0.5-\labelddist) {1};
	\node at (0.5-\labelddist,2.5+\labelddist) {3};
	\node at (0.5-\labelddist,0.5-\labelddist) {2};
	\node at (1.5+\labelddist,2.5+\labelddist) {4};
	\end{tikzpicture}
	=
	\begin{aligned}
	\begin{tikzpicture}[baseline={($(n1.base) + (0,1)$)},scale=0.8,rotate=90]
	%                          \draw (1,1) -- (1,2);
	\draw (0.5,0.5) -- (1,1);
	\draw (1.5,0.5) -- (1,1);
	\draw (1.5,1.5) -- (1,1);
	\draw (0.5,1.5) -- (1,1);
	\node[db] at (1,1) {};
	%                          \node[db] at (1,2) {};
	\node at (1.5+\labelddist,0.5-\labelddist) {4};
	\node at (0.5-\labelddist,1.5+\labelddist) {2};
	\node (n1) at (0.5-\labelddist,0.5-\labelddist) {1};
	\node at (1.5+\labelddist,1.5+\labelddist) {3};
	\end{tikzpicture}
	\end{aligned}
	=
	\begin{tikzpicture}[baseline={($(n1.base) + (0,0.6)$)},scale=0.8,rotate=90]
	\draw (1,1) -- (1,2);
	\draw (0.5,0.5) -- (1,1);
	\draw (1.5,0.5) -- (1,1);
	\draw (1.5,2.5) -- (1,2);
	\draw (0.5,2.5) -- (1,2);
	\node[db] at (1,1) {};
	\node[db] at (1,2) {};
	\node at (1.5+\labelddist,0.5-\labelddist) {4};
	\node at (0.5-\labelddist,2.5+\labelddist) {2};
	\node (n1) at (0.5-\labelddist,0.5-\labelddist) {1};
	\node at (1.5+\labelddist,2.5+\labelddist) {3};
	\end{tikzpicture}
	\\
	a) \hspace*{6cm} b) \\
	\begin{tikzpicture}[baseline={($(n1.base) + (0,0.6)$)},scale=0.8,rotate=90]
	%\draw (1,1) -- (1,2);
	\draw (0.5,0.5) -- (1,1);
	\draw (1.5,0.5) -- (1,1);
	\draw (1.5,2.5) -- (1,2);
	\draw (0.5,2.5) -- (1,2);
	\draw plot [smooth,tension=2] coordinates {(1,1) (1.2,1.5) (1,2)};
	\draw plot [smooth,tension=2] coordinates {(1,1) (0.8,1.5) (1,2)};
	\node[db] at (1,1) {};
	\node[dw] at (1,2) {};
	\node at (1.5+\labelddist,0.5-\labelddist) {4};
	\node at (0.5-\labelddist,2.5+\labelddist) {2};
	\node (n1) at (0.5-\labelddist,0.5-\labelddist) {1};
	\node at (1.5+\labelddist,2.5+\labelddist) {3};
	\end{tikzpicture}
	=
	\begin{tikzpicture}[baseline={($(n1.base) + (0,0.6)$)},scale=0.8,rotate=90]
	\draw (1,1) -- (1,2);
	\draw (0.5,0.5) -- (1,1);
	\draw (1.5,0.5) -- (1,1);
	\draw (1.5,2.5) -- (1,2);
	\draw (0.5,2.5) -- (1,2);
	\node[db] at (1,1) {};
	\node[dw] at (1,2) {};
	\node at (1.5+\labelddist,0.5-\labelddist) {4};
	\node at (0.5-\labelddist,2.5+\labelddist) {2};
	\node (n1) at (0.5-\labelddist,0.5-\labelddist) {1};
	\node at (1.5+\labelddist,2.5+\labelddist) {3};
	\end{tikzpicture} \\
	c)
	\end{gather*}
	\caption{Equivalence moves for on-shell diagrams: a) square move, b) merge/unmerge move for black nodes (similar for white nodes), c) bubble reduction.}
	\label{EquivalenceMoves}
\end{figure}

We are expecting the same property of regularity in the holomorphic soft limits with respect to variables parameterizing off-shell gluons  also in the case of gauge invariant amplitudes with several off-shell gluons (Wilson line insertions) $A^*_{k,m+n}$ (which is color ordered version of (\ref{AmplitudeSeveralOffShellGluons}) and $k$ is related to the total helicity $\lambda_{\Sigma}$ of
on-shell particles as $\lambda_{\Sigma}=m-2k+2n$). Namely, the amplitude should be regular in the limit $\xi_j \mapsto \epsilon\xi_j$, $\epsilon \rightarrow 0$, $j=1,\ldots,n$ where
$\xi_j$ is the spinor associated with the auxiliary vector $q_i$ from $k_T$ decomposition of $i$'th off-shell gluon momentum $k_i$. If the Grassmannian integral description is possible for such objects, then the Grassmannian integral should be regular in corresponding limits. Here we want to conjecture a representation for such integral using this regularity requirement. Moreover,
we want to describe the external kinematical data in a way similar to the case of one off-shell gluon. The last requirement suggests
that we should consider integral over $Gr(k,m+2n)$ Grassmannian, where $m$ is the number of on-shell momenta, and $n$ is the number of off-shell onces.

In the case of amplitudes with one off-shell gluon $A^*_{k,n+1}$ integrands of corresponding Grassmannian integrals (\ref{GrassmannianIntegralForOneOffShellGluon}) contained
products of delta functions of linear combinations of kinematical data and Grassmannian local coordinates (given by elements of $C$ matrix) together with the product of consecutive minors of $C$ matrix identical to the case of the Grassmannian integral $L^k_{n+2}[\Gamma]$:
\begin{equation}
\frac{\delta^{k\times 2} \left( C
\cdot \tilde{\lambda} \right)
\delta^{k\times 4} \left(C \cdot \tilde{\eta} \right)
\delta^{(n+2-k)\times 2} \left(C^{\perp} \cdot \lambda \right)}{(1 \cdots k)\cdots
(n+1\; n+2 \cdots k-2)(n+2 \; 1\cdots k-1)}.
\end{equation}
We want to use similar ingredients also in the case of $n$ off-shell gluons. In addition we need an insertion of some regulating function $Reg(m+1,\ldots,m+n).$ which should regulate holomorphic soft limit behavior with respect to $\xi_j$ variables. Thus, we are going to consider Grassmannian integral of the form
\begin{eqnarray}\label{GrassmannianIntegralForMultipleOffShellGluons}
\Omega_{m+2n}^k[\Gamma]&=&
\int_{\Gamma}\frac{d^{k\times (m+2n)}C}{\text{Vol}[GL (k)]}Reg.(m+1,\ldots,m+n)\times\nonumber\\
&\times&\frac{\delta^{k\times 2} \left( C
\cdot \vltuu \right)
\delta^{k\times 4} \left(C \cdot \vleuu \right)
\delta^{(m+2n-k)\times 2} \left(C^{\perp} \cdot \vlluu \right)}{(1 \cdots k)\cdots (m\cdots m+k-1) (m+1\cdots m+k)\cdots(m+2n \cdots k-1)},\nonumber\\
\end{eqnarray}
where external kinematical variables are chosen as
\begin{align}\label{SpinorsInDeformadGrassmannianMultipleOffShellMomenta}
&\vlluu_i = \vll_i, & i = 1,\ldots  m& ,
&\vlluu_{m+2j-1} &= \lambda_{p_j} ,
&\vlluu_{m+2j} &= \xi_j, & j = 1,\ldots  n, \nonumber \\
&\vltuu_i = \vlt_i, & i = 1,\ldots  m& ,
&\vltuu_{m+2j-1} &= \frac{\la\xi_j |k_{m+j}}{\la\xi_j p_j\ra},
&\vltuu_{m+2j} &= - \frac{\la p_j |k_{m+j}}{\la\xi_j p_j\ra} , & j = 1,\ldots  n, \nonumber \\
&\vleuu_i = \vlet_i,  & i = 1,\ldots  m& , &\vleuu_{m+2j-1} &= \vlet_{p_j} ,
&\vleuu_{m+2j} &= 0 , & j = 1,\ldots  n, \nonumber \\
\end{align}
and $Reg.(m+1,\ldots,m+n)$ is some function of local Grassmannian coordinates and external kinematical variables which should regulate the soft limits $\xi_j \mapsto \epsilon\xi_j$, $\epsilon \rightarrow 0$, $j=1,\ldots,n$. Based on our results in the $n=1$ case, we conjecture that $Reg.(m+1,\ldots,m+n)$ function should have factorized form\footnote{The numeration of columns in minors is understood up $\mbox{mod}(n+2m)$.} given by the product of $Reg.$ functions from $n=1$ case:
\begin{eqnarray}\label{RegFunctionNew}
&&Reg.(m+1,\ldots,m+n) = \prod_{j=1}^n Reg(j+m),\nonumber\\
&&Reg.(j+m) = \frac{\la\xi_j p_j\ra}{\kappa^{*}_j}\frac{(2j+m~~2j+1+m\cdots 2j+k-1+m)}{(2j-1+m~~2j+1+m\cdots 2j+k-1+m)}.
\end{eqnarray}
Direct evaluation of this Grassmannian integral for some explicit examples, presented in next section, show that indeed it is likely to be the correct representation for $A^*_{k,m+n}$. That is, we want to show that given an appropriate choice of integration contour $\Gamma=\Gamma_{tree}$ the following identity holds
\begin{equation}
A^*_{k,m+n}(\Omega_1,\ldots,\Omega_m,g^*_{m+1},\ldots,g^*_{m+n})=
\prod_{j=1}^n\frac{\partial^4}{\partial \tilde{\eta}_{p_j}^4}\Omega_{m+2n}^k[\Gamma_{tree}].
\end{equation}
Here, as before, $\Omega_i$ is an $i$-th on-shell $\mathcal{N}=4$ chiral superfield and  $g^*_j$ are off-shell gluons (Wilson line operator insertions). In addition, it is likely that $\Gamma_{tree}$ can be chosen identical to the case of $A_{k,m+2n}$ on-shell amplitude at least in some cases.
\begin{figure}[htbp]
	\begin{gather*}
	\begin{aligned}
	\begin{tikzpicture}[baseline={($(na.base) - (0,2)$)},transform shape,scale=1]
	\draw (0,0) -- (1,0);
	\draw (2.6,0) -- (3.6,0);
	\draw (0,-1) -- (3.6,-1);
	\draw (0,0) -- (0,-1);
	\draw (1,0) -- (1,-1);
	\draw (2.6,0) -- (2.6,-1);
	\draw (3.6,0) -- (3.6,-1);
	\draw (1.8,0.5) -- (1,0);
	\draw (1.8,0.5) -- (2.6,0);
	\draw (1.8,0.5) -- (1.8,1);
	\draw (0,0) -- (-0.5,0.3);
	\draw (0,-1) -- (-0.5,-1.3);
	\draw (3.6,0) -- (4.1,0.3);
	\draw (3.6,-1) -- (4.1,-1.3);
	\node[db] (na) at (0,0) {};
	\node[dw] at (1,0) {};
	\node[dw] at (2.6,0) {};
	\node[db] at (3.6,0) {};
	\node[dw] at (0,-1) {};
	\node[db] at (1,-1) {};
	\node[db] at (2.6,-1) {};
	\node[dw] at (3.6,-1) {};
	\node[dw] at (1.8,0.5) {};	
	\node at (1.8+\labelhdist,1) {$1$};	
	\node at (4 + \labelhdist,-0.5) {$2^*$};	
	\node at (-0.3 - \labelhdist,-0.5) {$3^*$};
	\end{tikzpicture}
	\end{aligned}
	\quad\longrightarrow\quad
	\begin{aligned}
	\begin{tikzpicture}[baseline={($(na.base) - (0,2)$)},transform shape, scale=1]
	\draw (0,0) -- (1,0);
	\draw (1,0) -- (2,0);
	\draw (0,-1) -- (3.6,-1);
	\draw (0,0) -- (0,-1);
	\draw (1,0) -- (1,-1);
	\draw (2,0) -- (2,-1);
	\draw (3.6,0) -- (3.6,-1);
	\draw (2.8,0.5) -- (2,0);
	\draw (2.8,0.5) -- (3.6,0);
	\draw (2.8,0.5) -- (2.8,1);
	\draw (0,0) -- (-0.5,0.3);
	\draw (0,-1) -- (-0.5,-1.3);
	\draw (3.6,0) -- (4.1,0.3);
	\draw (3.6,-1) -- (4.1,-1.3);
	\node[db] (na) at (0,0) {};
	\node[dw] at (1,0) {};
	\node[dw] at (2,0) {};
	\node[db] at (3.6,0) {};
	\node[dw] at (0,-1) {};
	\node[db] at (1,-1) {};
	\node[db] at (2,-1) {};
	\node[dw] at (3.6,-1) {};
	\node[dw] at (2.8,0.5) {};	
	\node at (2.8+\labelhdist,1) {$1$};	
	\node at (4 + \labelhdist,-0.5) {$2^*$};	
	\node at (-0.3 - \labelhdist,-0.5) {$3^*$};
	\end{tikzpicture}
	\end{aligned} \\
	\longrightarrow\quad
	\begin{aligned}
	\begin{tikzpicture}[baseline={($(na.base) - (0,2)$)},transform shape, scale=1]
	\draw (0,0) -- (1,0);
	\draw (0,-1) -- (2.6,-1);
	\draw (0,0) -- (0,-1);
	\draw (1,0) to[out=-45,in=45] (1,-1);
	\draw (1,0) to[out=-135,in=135] (1,-1);
	\draw (2.6,0) -- (2.6,-1);
	\draw (1.8,0.5) -- (1,0);
	\draw (1.8,0.5) -- (2.6,0);
	\draw (1.8,0.5) -- (1.8,1);
	\draw (0,0) -- (-0.5,0.3);
	\draw (0,-1) -- (-0.5,-1.3);
	\draw (2.6,0) -- (3.1,0.3);
	\draw (2.6,-1) -- (3.1,-1.3);
	\node[db] (na) at (0,0) {};
	\node[dw] at (1,0) {};
	\node[db] at (2.6,0) {};
	\node[dw] at (0,-1) {};
	\node[db] at (1,-1) {};
	\node[dw] at (2.6,-1) {};
	\node[dw] at (1.8,0.5) {};	
	\node at (1.8+\labelhdist,1) {$1$};	
	\node at (3 + \labelhdist,-0.5) {$2^*$};	
	\node at (-0.3 - \labelhdist,-0.5) {$3^*$};
	\end{tikzpicture}
	\end{aligned}\quad
	\longrightarrow\quad
	\begin{aligned}
	\begin{tikzpicture}[baseline={($(na.base) - (0,2)$)},transform shape, scale=1]
	\draw (0,0) -- (1,0);
	\draw (0,-1) -- (2.6,-1);
	\draw (0,0) -- (0,-1);
	\draw (1,0) -- (1,-1);
	\draw (2.6,0) -- (2.6,-1);
	\draw (1.8,0.5) -- (1,0);
	\draw (1.8,0.5) -- (2.6,0);
	\draw (1.8,0.5) -- (1.8,1);
	\draw (0,0) -- (-0.5,0.3);
	\draw (0,-1) -- (-0.5,-1.3);
	\draw (2.6,0) -- (3.1,0.3);
	\draw (2.6,-1) -- (3.1,-1.3);
	\node[db] (na) at (0,0) {};
	\node[dw] at (1,0) {};
	\node[db] at (2.6,0) {};
	\node[dw] at (0,-1) {};
	\node[db] at (1,-1) {};
	\node[dw] at (2.6,-1) {};
	\node[dw] at (1.8,0.5) {};	
	\node at (1.8+\labelhdist,1) {$1$};	
	\node at (3 + \labelhdist,-0.5) {$2^*$};	
	\node at (-0.3 - \labelhdist,-0.5) {$3^*$};
	\end{tikzpicture}
	\end{aligned}
	\end{gather*}
	\caption{On-shell diagram transformations for $A^{*}_{1+2}$
	}
	\label{fig: A1+2}
\end{figure}
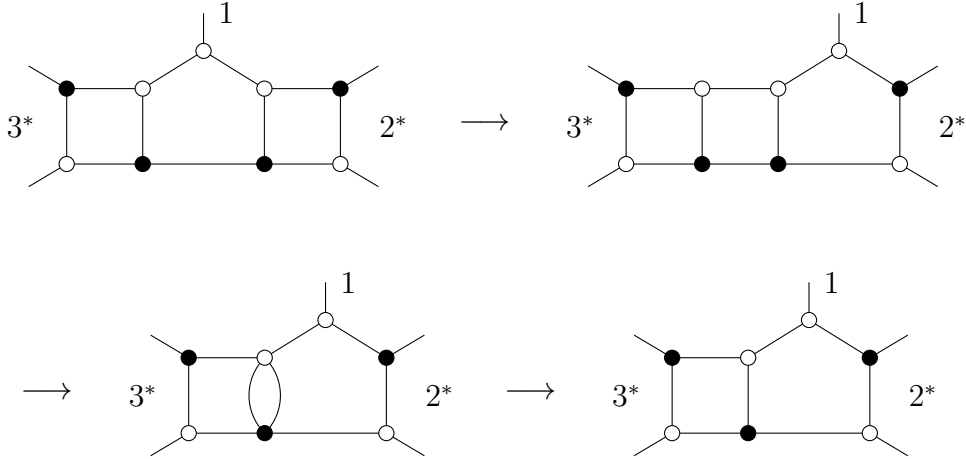

In \cite{offshell-1leg} we have also shown\footnote{See also the similar discussion for the case of form factors in \cite{SoftTheoremsFormFactors,FormFactorsGrassmanians,WilhelmThesis,q2zeroFormFactors} }, that on-shell diagrams for scattering amplitudes with one leg off-shell are given by corresponding on-shell diagrams for on-shell scattering amplitudes with one of the vertexes exchanged for off-shell vertex, see Fig. \ref{fig: offshell-vertex}. We expect that in the case of several off-shell legs the corresponding on-shell diagrams could be also obtained by similar procedure, that is exchanging on-shell vertexes containing off-shell legs with off-shell vertexes. Recalling our cutting and gluing procedure from \cite{offshell-1leg} it is easy to convince yourself that it is true in a case when off-shell legs are separated by on-shell ones. In a case when off-shell legs stay next to each other it is not as obvious. To see that it is actually the case, let us consider as example 3-point amplitude with two legs off-shell. In this case the corresponding on-shell diagram is the first diagram in Fig. \ref{fig: A1+2} times the corresponding $Reg.$ function (product of $Reg.$ functions from two off-shell vertexes). The latter could be taken out of integration sign\footnote{See appendix \ref{appA} for details.}, while the original off-shell diagram is reduced (see Fig. \ref{fig: A1+2}) to the on-shell diagram corresponding to our Grassmannian representation using equivalence moves from Fig. \ref{EquivalenceMoves}.

Finally, to end this section, we would like to stress the following feature of our conjecture. Namely, we may consider the situation when there are no external on-shell degrees of freedom at all ($m=0$). In this case we obtain the Grassmannian integral representation of color ordered correlation function of Wilson line operators (\ref{CorrFunctionOffShellGluons}):
\begin{equation}
A^*_{n}(g^*_{1},\ldots,g^*_{n})=
\prod_{j=1}^n\frac{\partial^4}{\partial \tilde{\eta}_{p_j}^4}\Omega_{2n}^n[\Gamma_{tree}].
\end{equation}

\section{Examples and checks}

Now we are going to reproduce results of BCFW recursion  \cite{vanHamerenBCFW1} for different off-shell amplitudes containing multiple off-shell gluons. We will start with calculations using spinor helicity representation and later see how similar computations could be performed using momentum twistor
representation.

\subsection{spinor helicity representation}

Let us first consider the simplest cases when Grassmannian integral fully localizes on delta functions. In the case of
$A^*_{k,1+2}(g^+_1,g^*_2,g^*_3)$ amplitude $k=2$ and the corresponding Grassmannian integral is given by
\begin{equation}
\Omega_{1+4}^2=
\int\frac{d^{2\times 5}C}{\text{Vol}[GL (2)]}Reg.(2)Reg.(3)
\frac{\delta^{2\times 2} \left( C
	\cdot \vltuu \right)
	\delta^{2\times 4} \left(C \cdot \vleuu \right)
	\delta^{3\times 2} \left(C^{\perp} \cdot \vlluu \right)}{(12)(23)(34)(45)(51)}.
\end{equation}
Here we integrate over $Gr(2,5)$ Grassmannian and the integral is fully localized on delta functions, so that the choice of integration contour prescription could be skipped. The $Reg.$ functions are given by ($m=1$)
\begin{equation}
Reg.(2)=\frac{\langle p_2\xi_2\rangle}{\kappa^{*}_2}\frac{(34)}{(24)},
~Reg.(3)=\frac{\langle p_3\xi_3\rangle}{\kappa^{*}_3}\frac{(51)}{(41)}.
\end{equation}
Solving delta function constraints we get
\begin{equation}
\prod_{j=2}^3\frac{\partial^4}{\partial \tilde{\eta}_{p_j}^4}\Omega_{1+4}^2=
\delta^4\left(\sum_{i=1}^{5}\vlluu_i\vltuu_i\right)\frac{\langle p_2\xi_2\rangle}{\kappa^{*}_2}\frac{\langle 34 \rangle}{\langle 24 \rangle}
\frac{\langle p_3\xi_3\rangle}{\kappa^{*}_3}\frac{\langle 15 \rangle}{\langle14\rangle}
\frac{\langle24\rangle^4}{\langle12\rangle\langle23\rangle\langle34\rangle\langle45\rangle\langle51\rangle}.
\end{equation}
Now, we should  comment on the conventions used for spinor products. First, we label all products of
$\epsilon_{\alpha\beta}\vlluu_i^{\alpha}\vlluu_j^{\beta}$ and $\epsilon_{\dot{\alpha}\dot{\beta}}\vltuu_i^{\dot{\alpha}}\vltuu_j^{\dot{\beta}}$ spinors as $\langle ij\rangle$ and $[ij]$. Next, to obtain final expressions we  need to use spinor redefinitions from (\ref{SpinorsInDeformadGrassmannianMultipleOffShellMomenta}). In the present case with $m=1$ and $n=2$: $\vlluu_1 = \vll_1$ and spinors $\vlluu_i$ with $i=2,\ldots,5$ are expressed in terms of $\xi_{1+j}$, $\lambda_{p_{1+j}}$, $j=1,2$. We will also use bra and ket notation
for spinors $\lambda_i \equiv |i\rangle$, $\tilde{\lambda}_i\equiv|i]$ sometime when it will make formulas more clear. Taking into account $k_T$ decomposition of off-shell momenta $k_2$ and $k_3$ the above expression for $\Omega_{1+4}^2$ is rewritten as (here and below $p_i=\lambda_i\tilde{\lambda}_i$, $p_i^2=0$, $k_i^2\neq 0$)
\begin{equation}
\prod_{j=2}^3\frac{\partial^4}{\partial \tilde{\eta}_{p_j}^4}\Omega_{1+4}^2=
\delta^4(p_1+k_2+k_3)\frac{\langle p_2\xi_2\rangle\langle p_3\xi_3\rangle}{\kappa^{*}_2\kappa^{*}_3}
\frac{\langle p_2p_3\rangle^4}{\langle1p_2\rangle\langle p_2\xi_2\rangle\langle p_2p_3\rangle\langle p_3\xi_3\rangle\langle1p_3\rangle},
\end{equation}
which is exactly the result of BCFW recursion from \cite{vanHamerenBCFW1}:
\begin{equation}
A^*_{2,1+2}(g^+,g^*_2,g^*_3)=
\delta^4(p_1+k_2+k_3)\frac{1}{\kappa^{*}_2\kappa^{*}_3}
\frac{\langle p_2p_3\rangle^3}{\langle1p_2\rangle\langle p_31\rangle}.
\end{equation}
As we already said before, the off-shell gluons could be actually  arbitrary distributed among on-shell ones. The general formula for such configurations will look rather complicated, but particular examples are not. As an example, let us reproduce known answer for
$$
A^*_{k,n+2}(g^*_1,g^+_2,\ldots,g_{i-2}^+,g^*_{i-1},g_{i}^+,\ldots,g^+_n)
$$
amplitude. As before $k=2$ and we are integrating over $Gr(2,n+4)$ Grassmannian. In this case the integral is also
localized on delta functions (note that in this particular example we use different labels for $Reg.$ functions compared to other examples since the positions of off-shell gluons are different) and is given by
\begin{equation}
\Omega_{n+4}^2=
\int\frac{d^{2\times (n+4)}C}{\text{Vol}[GL (2)]}Reg.(1)Reg.(i-1)
\frac{\delta^{2\times 2} \left( C
	\cdot \vltuu \right)
	\delta^{2\times 4} \left(C \cdot \vleuu \right)
	\delta^{(n+2)\times 2} \left(C^{\perp} \cdot \vlluu \right)}{(12)(23)\ldots(ii+1)(i+1i+2)\ldots(n+41)},
\end{equation}
with
\begin{equation}
Reg(1).=\frac{\langle p_1\xi_1\rangle}{\kappa^{*}_1}\frac{(23)}{(13)},
~Reg(i-1).=\frac{\langle p_{i-1}\xi_{i-1}\rangle}{\kappa^{*}_{i-1}}\frac{(i+1i+2)}{(ii+2)}.
\end{equation}
Evaluating the above integral we get (for saving space we skip
momentum conservation delta function
$\delta^4(k_1+p_2+\ldots+p_{i-2}+k_{i-1}+p_{i}+\ldots+p_n)$)
\begin{equation}
\prod_{j=1,i}\frac{\partial^4}{\partial \tilde{\eta}_{p_j}^4}\Omega_{n+4}^2=
\frac{\langle p_1\xi_1\rangle}{\kappa^{*}_1}\frac{\langle 23\rangle}{\langle 13\rangle}
\frac{\langle p_{i-1}\xi_{i-1}\rangle}{\kappa^{*}_{i-1}}\frac{\langle i+1 i+2 \rangle}{\langle i i+2 \rangle}
\frac{\langle 1i\rangle^4}{\langle12\rangle\langle 23\rangle\ldots\langle ii+1\rangle\langle i+1i+2\rangle\ldots\langle n+41\rangle},
\end{equation}
which after relabeling spinor variables
\begin{eqnarray}
\begin{matrix}
\vlluu_1&\vlluu_2&\vlluu_3&\ldots& \vlluu_{i-1}&\vlluu_i&\vlluu_{i+1}&\vlluu_{i+2}&\ldots&\vlluu_{n+4}\\
\downarrow&\downarrow&\downarrow&\downarrow&\downarrow&\downarrow&\downarrow&\downarrow&\downarrow&\downarrow\\
\lambda_{p_1}&\xi_1&\lambda_2&\ldots& \lambda_{i-2}&\lambda_{p_{i-1}}&\xi_{i-1}&\lambda_{i}&\ldots&\lambda_{n}
\end{matrix}
\end{eqnarray}
can be rewritten as
\begin{equation}
\prod_{j=1,i}\frac{\partial^4}{\partial \tilde{\eta}_{p_j}^4}\Omega_{n+4}^2=
\frac{1}{\kappa^{*}_1\kappa^{*}_{i-1}}
\frac{\langle p_{1}p_{i-1}\rangle^4}{\langle p_12\rangle\langle 23\rangle\ldots\langle i-2p_{i-1}\rangle\langle p_{i-1}i\rangle\ldots\langle np_1\rangle},
\end{equation}
The latter result is in agreement with the result of BCFW recursion from \cite{vanHamerenBCFW1}
\begin{equation}
A^*_{2,n+2}(g^*_1,g^+_2,\ldots,g_{i-2}^+,g^*_{i-1},g_{i}^+,\ldots,g^+_n)=
\frac{1}{\kappa^{*}_1\kappa^{*}_{i-1}}
\frac{\langle p_{1}p_{i-1}\rangle^4}{\langle p_12\rangle\langle 23\rangle\ldots\langle i-2p_{i-1}\rangle\langle p_{i-1}i\rangle\ldots\langle np_1\rangle}.
\end{equation}

Form this example it should be now clear what steps should be performed to include extra off-shell gluons (Wilson line operators insertions) in the presented Grassmannian integral representation
for off-shell amplitudes. Namely, one have to consider Grassmannian integral representation for the $m+2n$ point on-shell amplitude $L^k_{m+2n}[\Gamma]$ ($m$ and $n$ are numbers of on-shell
and off-shell states we are interested in). Next, one have to choose a pair of consecutive minors $(i~i+1\ldots,i+k-1)$ and $(i+1~i+2,\ldots,i+k)$. and add to the integrand of $L^k_{m+2n}[\Gamma]$ integral factor
\begin{equation}\label{RegFunctionNew}
\frac{\la\xi_j p_j\ra}{\kappa^{*}_j}\frac{(i+1~i+2,\ldots,i+k)}{(i~i+2,\ldots,i+k)},
\end{equation}
together with replacement of spinors $\lambda_l,\tilde{\lambda}_l,\tilde{\eta}_l$, $l=i,i+1$ with
\begin{eqnarray}
\begin{matrix}
\lambda_i&\lambda_{i+1}&\tilde{\lambda}_i&\tilde{\lambda}_{i+1}&\tilde{\eta}_i&\tilde{\eta}_{i+1}\\
\downarrow&\downarrow&\downarrow&\downarrow&\downarrow&\downarrow\\
\lambda_{p_j}&\xi_j&\frac{\la\xi_j |k_{j}}{\la\xi_j p_j\ra}&-\frac{\la p_j |k_{j}}{\la\xi_j p_j\ra}&\tilde{\eta}_{p_j}&0
\end{matrix}
\end{eqnarray}
Here $k_j$ is the momentum for the $j$-th off-shell gluon and the label $j$ should be chosen in such a way that the consecutive numeration of particle momenta is restored. All other spinor labels should be relabeled accordingly. To obtain $n$ off-shell gluon insertions this operation should be repeated $n-1$ times. This way (\ref{GrassmannianIntegralForMultipleOffShellGluons}) shows the result of these steps when off-shell gluons are inserted "one after another".

Next, let us consider $A^*_{k,1+2}(g^-_1,g^*_2,g^*_3)$ amplitude. In this case $k=3$ and the integral over $Gr(3,5)$ Grassmannian is again localized on delta functions:
\begin{equation}
\Omega_{1+4}^3=
\int\frac{d^{3\times 5}C}{\text{Vol}[GL (3)]}Reg.(2)Reg.(3)
\frac{\delta^{3\times 2} \left( C
	\cdot \vltuu \right)
	\delta^{3\times 4} \left(C \cdot \vleuu \right)
	\delta^{2\times 2} \left(C^{\perp} \cdot \vlluu \right)}{(123)(234)(345)(451)(512)}.
\end{equation}
The $Reg.$ functions are given by ($m=1$):
\begin{equation}
Reg.(2)=\frac{\langle p_2\xi_2\rangle}{\kappa^{*}_2}\frac{(345)}{(245)},
~Reg.(3)=\frac{\langle p_3\xi_3\rangle}{\kappa^{*}_3}\frac{(512)}{(412)}.
\end{equation}
Evaluating integral we get
\begin{equation}
\frac{\partial^4}{\partial \tilde{\eta}_{1}^4}\prod_{j=2}^3\frac{\partial^4}{\partial \tilde{\eta}_{p_j}^4}\Omega_{1+4}^3=
\delta^4\left(\sum_{i=1}^{5}\vlluu_i\vltuu_i\right)\frac{\langle p_2\xi_2\rangle}{\kappa^{*}_2}\frac{[ 12 ]}{[ 13 ]}
\frac{\langle p_3\xi_3\rangle}{\kappa^{*}_3}\frac{[ 34 ]}{[35]}
\frac{[35]^4}{[12][23][34][45][51]}.
\end{equation}
Expressing spinors in terms of external kinematical data using (\ref{SpinorsInDeformadGrassmannianMultipleOffShellMomenta})
with $m=1$ and $n=2$ together with $k_T$ decomposition of  off-shell momenta $k_i$ ($q_i=\xi_i\tilde{\xi}_i$, $i=2,3$):
%\begin{equation}
%
%\vltuu_3=\frac{k_2|p_2\rangle}{\langle p_2\xi_2\rangle}=\frac{\kappa^{*}_2}{\langle p_2\xi_2\rangle}\tilde{\lambda}_{p_2},~\vltuu_5=\frac{k_3|p_3\rangle}{\langle p_3\xi_3\rangle}=\frac{\kappa^{*}_3}{\langle p_3\xi_3\rangle}\tilde{\lambda}_{p_3},
%
%~k_i^2=\kappa_i\kappa^{*}_i,~i=2,3,
%
%\end{equation}
%
\begin{equation}
\vltuu_3=\frac{k_2|p_2\rangle}{\langle p_2\xi_2\rangle}=\frac{\kappa^{*}_2}{\langle p_2\xi_2\rangle}|p_2],~\vltuu_5=\frac{k_3|p_3\rangle}{\langle p_3\xi_3\rangle}=\frac{\kappa^{*}_3}{\langle p_3\xi_3\rangle}|p_3],
~k_i^2=\kappa_i\kappa^{*}_i,~i=2,3,
\end{equation}
and similar expressions for $\vltuu_2,\vltuu_4$ we get (here we again dropped momentum conservation delta function $\delta^4(p_1+k_2+k_3)$):
\begin{eqnarray}
\frac{\partial^4}{\partial \tilde{\eta}_{1}^4}\prod_{j=2}^3\frac{\partial^4}{\partial \tilde{\eta}_{p_j}^4}\Omega_{1+4}^3&=&
\frac{\langle p_2\xi_2\rangle\langle p_3\xi_3\rangle}{\kappa^{*}_2\kappa^{*}_3}
\frac{[p_2p_3]^3\kappa^{*3}_2\kappa^{*3}_3\langle p_2\xi_2\rangle^{-3}\langle p_3\xi_3\rangle^{-3}}{[1p_2]\kappa^{*}_2\langle p_2\xi_2\rangle^{-1}
	\kappa_2\kappa^{*}_2\langle p_2\xi_2\rangle^{-1}\kappa_3\kappa^{*}_3\langle p_3\xi_3\rangle^{-1}[p_31]\kappa^{*}_3\langle p_3\xi_3\rangle^{-1}}\nonumber\\
&=&\frac{1}{\kappa_2\kappa_3}\frac{[p_2p_3]^3}{[1p_2][p_31]}.
\end{eqnarray}
This is exactly the result of BCFW recursion  \cite{vanHamerenBCFW1} for $A^*_{3,1+2}(g^-_1,g^*_2,g^*_3)$ amplitude:
\begin{eqnarray}
A^*_{3,1+2}(g^-_1,g^*_2,g^*_3)
&=&\delta^4(p_1+k_2+k_3)\frac{1}{\kappa_2\kappa_3}\frac{[p_2p_3]^3}{[1p_2][p_31]}.
\end{eqnarray}

It is interesting to consider amplitudes with only off-shell states present ($m=0$). The simplest example of this kind is given by $A^*_{k,0+3}(g_1^*,g_2^*,g_3^*)$ amplitude. In this case $k=3$ and integration goes over $Gr(3,6)$ Grassmannian. The Grassmannian integral in this case is no longer trivial and does not localizes on delta functions. It can be however reduced to the integral over one complex parameter $\tau$, which in its turn could be evaluated by taking residues (see \cite{DualitySMatrix,offshell-1leg}). The result of BCFW recursion for this amplitude is given by \cite{vanHamerenBCFW1}:
\begin{eqnarray}
A^*_{3,0+3}(g_1^*,g_2^*,g_3^*)=\delta^4(k_1+k_2+k_3)(1+\mathbb{P}'+\mathbb{P}'^2)
\frac{1}{\kappa_1^*\kappa_3}
\frac{\langle p_1p_2 \rangle^3 [p_2p_3]^3}{\langle p_2|k_1|p_2] \langle p_2|k_1|p_3] \langle p_1|k_3|p_2]},\nonumber\\
\end{eqnarray}
where $\mathbb{P}'$ is the permutation operator shifting all spinor and momenta labels by $+1$ $\mbox{mod}(3)$. It should be stressed that this object can be considered as a correlation function of three gauge invariant operators given by Wilson lines. Now, let us proceed with the Grassmannian integral itself. The conjectured Grassmannian integral representation in this case is given by the following integral
\begin{equation}
\Omega_{0+6}^3[\Gamma]=
\int_{\Gamma}\frac{d^{3\times 6}C}{\text{Vol}[GL (3)]}\prod_{i=1}^3Reg.(i)
\frac{\delta^{3\times 2} \left( C
	\cdot \vltuu \right)
	\delta^{3\times 4} \left(C \cdot \vleuu \right)
	\delta^{3\times 2} \left(C^{\perp} \cdot \vlluu \right)}{(123)(234)(345)(456)(561)(612)},
\end{equation}
where
\begin{equation}
Reg.(1)=\frac{\langle p_1\xi_1\rangle}{\kappa^{*}_1}\frac{(234)}{(134)},
~Reg.(2)=\frac{\langle p_2\xi_2\rangle}{\kappa^{*}_2}\frac{(456)}{(356)},
~Reg.(3)=\frac{\langle p_3\xi_3\rangle}{\kappa^{*}_3}\frac{(612)}{(512)}.
\end{equation}

As we already mentioned this integral can be reduced to the integral over single complex parameter $\tau$. Next, we fix $GL(3)$ gauge as in \cite{Henrietta_Amplitudes} in the case of $\mbox{NMHV}_6$ amplitude. The minors $(123)$, $(345)$, $(561)$ in this case will became linear functions of parameter $\tau$ and
we choose integration contour identical to the case of $\mbox{NMHV}_6$ amplitude. This choice means, that we are interested in residues at the zeros of minors $(123)$, $(345)$ and $(561)$, which we will label $\{1\}$, $\{3\}$, $\{5\}$. The corresponding integration contour $\Gamma=\Gamma_{135}$ just encircles these poles and we get
\begin{equation}
\prod_{j=1,3,5}^3\frac{\partial^4}{\partial \vleuu_j^4}~\Omega_{0+6}^3[\Gamma_{135}]=\{1\}+\{3\}+\{5\}.
\end{equation}
In the case of $\{1\}$ residue the corresponding minors are given by
\begin{eqnarray}
(123)(\tau)&=&\tau\langle13\rangle[46],
~(512)|_{\tau=0}=\frac{\langle12\rangle}{\langle13\rangle},
~(134)|_{\tau=0}=\frac{[56]}{[46]},\nonumber\\
(356)|_{\tau=0}&=&\frac{\langle3|1+2|4]}{\langle13\rangle[46]},
~(561)|_{\tau=0}=\frac{\langle1|2+3|4]}{\langle13\rangle[46]},
~(345)|_{\tau=0}=\frac{\langle3|1+2|6]}{\langle13\rangle[46]},\nonumber\\
\end{eqnarray}
and for the residue $\{1\}$ itself we get
\begin{eqnarray}
\{1\}&=&\delta^4(k_1+k_2+k_3)
\frac{\langle p_1\xi_1\rangle\langle p_2\xi_2\rangle\langle p_3\xi_3\rangle}{\kappa^{*}_1\kappa^{*}_2\kappa^{*}_3}\frac{1}{\langle13\rangle[46]}
\frac{1}{(134)(345)(356)(561)(512)}\Big{|}_{\tau=0}\nonumber\\
&=&\delta^4(k_1+k_2+k_3)\frac{\langle p_1\xi_1\rangle\langle p_2\xi_2\rangle\langle p_3\xi_3\rangle}{\kappa^{*}_1\kappa^{*}_2\kappa^{*}_3}
\frac{\langle13\rangle^3[46]^3}{\langle12\rangle[56]\langle3|1+2|4]\langle3|1+2|6]\langle1|5+6|4]},
\nonumber\\
\end{eqnarray}
where we used the definition of external kinematical variables from (\ref{SpinorsInDeformadGrassmannianMultipleOffShellMomenta}) with $m=0$ in the argument of momentum conservation delta function. The other  residues can be obtained by the action of permutation operator $\mathbb{P}$ shifting labels of $\vlluu_i$ and $\vltuu_i$ spinors by $+1$ $\mbox{mod}(6)$
\begin{eqnarray}
\{3\}=\delta^4(k_1+k_2+k_3)\frac{\langle p_1\xi_1\rangle\langle p_2\xi_2\rangle\langle p_3\xi_3\rangle}{\kappa^{*}_1\kappa^{*}_2\kappa^{*}_3}
\mathbb{P}^2\frac{\langle13\rangle^3[46]^3}{\langle12\rangle[56]\langle3|1+2|4]\langle3|1+2|6]\langle1|5+6|4]},
\nonumber\\
\end{eqnarray}
\begin{eqnarray}
\{5\}=\delta^4(k_1+k_2+k_3)\frac{\langle p_1\xi_1\rangle\langle p_2\xi_2\rangle\langle p_3\xi_3\rangle}{\kappa^{*}_1\kappa^{*}_2\kappa^{*}_3}
\mathbb{P}^4\frac{\langle13\rangle^3[46]^3}{\langle12\rangle[56]\langle3|1+2|4]\langle3|1+2|6]\langle1|5+6|4]}.
\nonumber\\
\end{eqnarray}
Using the definition of external kinematical variables (\ref{SpinorsInDeformadGrassmannianMultipleOffShellMomenta}) in other parts of expression for $\{1\}$ residue and dropping for brevity overall momentum conservation delta function we get
\begin{eqnarray}
\{1\}&=&\frac{\langle p_1\xi_1\rangle\langle p_2\xi_2\rangle\langle p_3\xi_3\rangle}{\kappa^{*}_1\kappa^{*}_2\kappa^{*}_3}
\frac{\langle p_1p_2\rangle^3[p_2p_3]^3\kappa^{*3}_2\kappa^{*3}_3\langle p_2\xi_2\rangle^{-3}\langle p_3\xi_3\rangle^{-3}}
{\kappa_3\kappa^{*}_3\langle p_3\xi_3\rangle^{-2}\langle p_1\xi_1\rangle\kappa^{*}_3\kappa^{*2}_2
	\langle p_2\xi_2\rangle^{-2}
	\langle p_2|k_1|p_2] \langle p_2|k_1|p_3] \langle p_1|k_3|p_2]}
\nonumber\\
&=&\frac{1}{\kappa_1^*\kappa_3}
\frac{\langle p_1p_2 \rangle^3 [p_2p_3]^3}{\langle p_2|k_1|p_2] \langle p_2|k_1|p_3] \langle p_1|k_3|p_2]}.
\end{eqnarray}
The other residues are then given by
\begin{eqnarray}
\{3\}=\mathbb{P}'\{1\},~\{5\}=\mathbb{P}'^2\{1\}.
\end{eqnarray}
So, we see that indeed the following relation holds
\begin{eqnarray}
\prod_{j=1,2,3}^3\frac{\partial^4}{\partial \tilde{\eta}_{p_j}^4}~\Omega_{0+6}^3[\Gamma_{135}]=\{1\}+\{3\}+\{5\}=A^*_{3,0+3}(g_1^*,g_2^*,g_3^*),
\end{eqnarray}
and our conjectured Grassmannian integral correctly reproduces 3-point amplitude with three off-shell gluons or equivalently color ordered correlation function of three Wilson line operators:
\begin{eqnarray}\label{CorrFunctionOffthreeShellGluons}
\la 0|\mathcal{W}_{p_{1}}^{c_{1}}(k_{1})\mathcal{W}_{p_{2}}^{c_{2}}(k_{2})\mathcal{W}_{p_{3}}^{c_{3}}(k_{3})|0\ra^{tree}&=&g~tr(t^{c_1}t^{c_2}t^{c_3})A^*_{3,0+3}(g_1^*,g_2^*,g_3^*)
\nonumber\\
&+&g~tr(t^{c_1}t^{c_3}t^{c_2})A^*_{3,0+3}(g_1^*,g_3^*,g_2^*).
\end{eqnarray}
%%MARK 111

Now let us consider two 4-point amplitudes with two off-shell gluons, which we will need when discussing the vacuums for amplitudes with two off-shell gluons in the context of the  auxiliary $\mathfrak{gl}(4|4)$ spin chain in the next section. The Grassmannian integral representation for $A^*_{2,2+2}(1^*,2^*,3^+,4^+)$ is given by
\begin{equation}
	\Omega_{2+4}^2=
	\int_{\Gamma}\frac{d^{2\times 6}C}{\text{Vol}[GL (2)]}\prod_{i=1}^2Reg.(i)
	\frac{\delta^{2\times 2} \left( C
		\cdot \vltuu \right)
		\delta^{2\times 4} \left(C \cdot \vleuu \right)
		\delta^{4\times 2} \left(C^{\perp} \cdot \vlluu \right)}{(12)(23)(34)(45)(56)(61)},
\end{equation}
where
\begin{equation}
	Reg.(1)=\frac{\langle p_1\xi_1\rangle}{\kappa^{*}_1}\frac{(23)}{(13)},
	~Reg.(2)=\frac{\langle p_2\xi_2\rangle}{\kappa^{*}_2}\frac{(45)}{(35)}.
\end{equation}
Fixing $GL(2)$ gauge as
\begin{eqnarray}
	C=\left( \begin{array}{cccccc}
		1 & 0 & c_{13} & c_{14} & c_{15} & c_{16} \\
		0 & 1 & c_{23} & c_{24} & c_{25} & c_{26}
  \end{array} \right).
\end{eqnarray}
and solving delta functions constraints we get
\begin{eqnarray}
	c_{1i}=\frac{\la 2 i\ra}{\la 2 1\ra},\quad  c_{2i}=\frac{\la 1 i\ra}{\la 1 2\ra} .
\end{eqnarray}
Performing required spinor substitutions
\begin{eqnarray}
\vlluu_1 = |p_1\rangle, \vlluu_2  = |\xi_1\rangle,
\vlluu_3 = |p_2\rangle, \vlluu_4  = |\xi_2\rangle,
\vlluu_5 = |3\rangle, \vlluu_6 = |4 \rangle.
\end{eqnarray}
%\begin{eqnarray}
%
%\vlluu_1 = \vll_{p_1}, \vlluu_2  = \vll_{\xi_1},
%
%\vlluu_3 = \vll_{p_2}, \vlluu_4  = \vll_{\xi_2},
%
%\vlluu_5 = \vll_3, \vlluu_6 = \vll_4 .
%
%\end{eqnarray}
%
the considered off-shell amplitude is given by
\begin{eqnarray}
A^*_{2,2+2}(1^*,2^*,3^+,4^+)&=&\prod_{j=1,3}\frac{\partial^4}{\partial \vleuu_j^4}~\Omega_{2+4}^2= \frac{1}{\kappa_1^*\kappa_2^*}
\frac{\la p_1 p_2\ra^3}{\la p_2 3\ra\la 3 4\ra\la 4 p_1\ra}\delta^4 (k_1 + k_2 + p_3 + p_4). \nonumber \\ \label{A*(1*,2*,3+,4+)}
\end{eqnarray}
in total agreement with the result of BCFW recursion \cite{vanHamerenBCFW1}. The consideration of $A^*(1^*,2^*,3^-,4^-)$ amplitude is similar. The Grassmannian integral representation in this case is given by
\begin{equation}
	\Omega_{2+4}^4=
	\int_{\Gamma}\frac{d^{4\times 6}C}{\text{Vol}[GL (4)]}\prod_{i=1}^2Reg.(i)
	\frac{\delta^{4\times 2} \left( C
		\cdot \vltuu \right)
		\delta^{4\times 4} \left(C \cdot \vleuu \right)
		\delta^{2\times 2} \left(C^{\perp} \cdot \vlluu \right)}{(1234)(2345)(3456)(4561)(5612)(6123)},
\end{equation}
where
\begin{equation}
Reg.(1)=\frac{\langle p_1\xi_1\rangle}{\kappa^{*}_1}\frac{(2345)}{(1345)},
~Reg.(2)=\frac{\langle p_2\xi_2\rangle}{\kappa^{*}_2}\frac{(4561)}{(3561)}.
\end{equation}
and $C$-matrix after gauge fixing
\begin{eqnarray}
C=\left( \begin{array}{cccccc}
1 & 0 & 0 & 0 & c_{15} & c_{16} \\
0 & 1 & 0 & 0 & c_{25} & c_{26} \\
0 & 0 & 1 & 0 & c_{35} & c_{36} \\
0 & 0 & 0 & 1 & c_{45} & c_{46}  \end{array} \right).
\end{eqnarray}
Solving delta functions constraints
\begin{eqnarray}
&&c_{15}=\frac{[16]}{[56]},  c_{16}=\frac{[15]}{[56]},  c_{25}=\frac{[26]}{[56]},  c_{26}=\frac{[25]}{[56]}, \nonumber\\
&&c_{35}=\frac{[36]}{[56]},  c_{36}=\frac{[35]}{[56]}, c_{45}=\frac{[46]}{[56]},  c_{46}=\frac{[45]}{[56]}.
\end{eqnarray}
and performing required spinor substitutions
\begin{eqnarray}
\vltuu_1=\frac{k_1|\xi_1\rangle}{c_1},
\vltuu_2=\frac{k_1|p_1\rangle}{c_1}=|p_1]\frac{\kappa_1^*}{c_1},
\vltuu_3=\frac{k_2|\xi_2\rangle}{c_1},
\vltuu_4=\frac{k_2|p_2\rangle}{c_2}=|p_2]\frac{\kappa_2^*}{c_2},
\vltuu_5=|3],\vltuu_6=|4],\nonumber\\
\end{eqnarray}
with $c_i \equiv \langle p_i\xi_i\rangle$ and $k_i^2=\kappa_i\kappa^*_i$, we get (dropping obvious momentum conservation delta function):
\begin{eqnarray}
	A^*_{4,2+2}(1^*,2^*,3^-,4^-)&=&\prod_{j=1,3,5,6}\frac{\partial^4}{\partial \vleuu_j^4}~\Omega_{2+4}^4=
	\frac{c_1}{\kappa_1^*}\frac{[16]}{[26]}\frac{c_2}{\kappa_2^*}\frac{[23]}{[24]}
	\frac{[24]^4}{[12][23][34][45][56][61]}\nonumber\\
	&=&\frac{c_1}{\kappa_1^*}\frac{c_2}{\kappa_2^*}\frac{[p_1p_2]^4c_2^{-4}c_1^{-4}\kappa_1^{*4}\kappa_2^{*4}}
	{\kappa_1\kappa_1^*c_1^{-1}c_1^{-1}c_2^{-1}\kappa_1^*\kappa_2^*\kappa_2\kappa_2^*c_2^{-1}c_2^{-1}
		c_1^{-1}\kappa_1^*\kappa_2^*[p_1p_2][p_23][34][4p_1]}\nonumber\\
	&=&
	\frac{1}{\kappa_1\kappa_2}\frac{[p_1p_2]^4}{[p_1p_2][p_23][34][4p_1]}. \label{A*(1*,2*,3-,4-)}
\end{eqnarray}
This result is again in agreement with BCFW recursion \cite{vanHamerenBCFW1}.

\subsection{momentum twistor representation}

Using the results\footnote{We are referring the reader to \cite{offshell-1leg} for the momentum twistor notation used here.} of \cite{offshell-1leg} and the discussion in previous section it is easy to see, that in the momentum twistor space the Grassmannian integral representation for amplitudes with two off-shell gluons takes the following form\footnote{The generalization to the cases with more off-shell legs is straightforward, also note that
$\Omega_{m+4}^2=A^*_{2,m+2}$.}:
\begin{eqnarray}
A^{*}_{k,m+2} &=& \prod_{i=m+1,m+3}\frac{\partial^4}{\partial\vleuu^4_i}\Big(~\Omega_{m+4}^2~\omega^k_{m+4}[\Gamma_{tree}]~\Big), \nonumber \\
\omega^k_{m+4}[\Gamma] &=& \int_{\Gamma}
\frac{d^{(k-2)\times (m+4)}D}{\text{Vol}[GL(k-2)]}Reg.(m+1)Reg.(m+2)
\frac{\delta^{ 4 (k-2) | 4 (k-2)} (D\cdot \mathcal{Z})}{(1 \; \ldots \; k-2) \;\ldots \; (m+4 \; \ldots \; k-3)}. \nonumber\\
\label{GrassmannianMomentumTwistors}
\end{eqnarray}
where the numeration of columns goes up to $\mbox{mod}(m+4)$) and $Reg.$ functions are given by the following expressions
\begin{eqnarray}
Reg.(m+2)&=&\frac{1}{1+\frac{\la p_{m+2}\xi_{m+2}\ra}{\la p_{m+2} 1\ra}\frac{(m+4 \ldots m+4+k-3)}{(1  \ldots  1+k-3)}},\\
Reg.(m+1)&=&\frac{1}{1+\frac{\la p_{m+1}\xi_{m+1}\ra}{\la p_{m+1} p_{m+2}\ra}
	\frac{(m+2 \ldots m+2+k-3)}{(m+3 \ldots m+3+k-3)}}.
\end{eqnarray}
In the case of $k=2$ the matrix $D$ is zero dimensional, all its consecutive minors equal to one and nonconsecutive to zero. So, the integral in (\ref{GrassmannianMomentumTwistors}) is zero dimensional, integrand equal to 1 and the result is given by $A^*_{2,m+4}$.

For the $k=3$ case we have
\begin{equation}
\omega^3_{m+4}[\Gamma]= \int_{\Gamma}\frac{d^{1\times (m+4)} D}{\text{Vol}[GL(1)]}
Reg.(m+1)Reg.(m+2)
\frac{1}{d_1\ldots d_m d_{m+1}\ldots d_{m+4}}\delta^{4|4} (D\cdot \mathcal{Z}),
\end{equation}
where
\begin{equation}
Reg.(m+2)=\frac{1}{1+\frac{\la p_{m+2}\xi_{m+2}\ra}{\la p_{m+2} 1\ra}\frac{d_{m+4}}{d_1}},~
Reg.(m+1)=\frac{1}{1+\frac{\la p_{m+1}\xi_{m+1}\ra}{\la p_{m+1} p_{m+2}\ra}
	\frac{d_{m+2}}{d_{m+3}}}.
\end{equation}
and the amplitude is presumably given by
\begin{eqnarray}\label{2OffShellgluonsNMHVamplitude}
A^*_{3,m+2}(\Omega_1,\ldots,\Omega_m,g_{m+1}^*,g_{m+2}^*) &\stackrel{?}{=}&
\prod_{i=m+1,m+3}\frac{\partial^4}{\partial\vleuu^4_i}\Big(~\Omega_{m+4}^2~
\omega^3_{m+4}[\Gamma_{tree}]~\Big) \nonumber \\ &=&
\prod_{i=m+1,m+3}\frac{\partial^4}{\partial\vleuu^4_i}\left(\Omega_{m+4}^2\left\{
\sum_{i <j}c_{ij}[1 i-1ij-1j]
\right\}\right). \nonumber \\
\end{eqnarray}
Here $\Gamma_{tree}$ is the $[1,2\ra$ contour for $A_{3,m+4}$ on-shell amplitude and for $c_{ij}$ coefficients we get
\begin{equation}
c_{ij}=0,~ \mbox{if there is no} ~m+3~ \mbox{label among} ~i-1,i,j-1,j.
\end{equation}
in other cases
\begin{eqnarray}
c_{m+3j}&=&0,\\
c_{i,m+3}&=&\frac{1}{1+\frac{\la p_{m+1}\xi_{m+1}\ra}{\la p_{m+1} p_{m+2}\ra}
	\frac{\la 1i-1im+3 \ra}{\la 1i-1im+2 \ra}},\\
c_{i,m+4}&=&\frac{1}{1+\frac{\la p_{m+2}\xi_{m+2}\ra}{\la p_{m+2} 1\ra}\frac{\la 1i-1im+3 \ra}{\la i-1im+3m+4 \ra}},\\
c_{m+2,m+4}&=&\frac{1}{1+\frac{\la p_{m+1}\xi_{m+1}\ra}{\la p_{m+1} p_{m+2}\ra}
	\frac{\la 1 m+1 m+3 m+4\ra}{\la 1 m+1 m+2 m+4 \ra}}
\frac{1}{1+\frac{\la p_{m+2}\xi_{m+2}\ra}{\la p_{m+2} 1\ra}\frac{\la 1 m+1 m+2 m+3 \ra}{\la m+1 m+2 m+3 m+4 \ra}}.
\end{eqnarray}

Now let us proceed with the particular examples of 4-point amplitudes with two off-shell gluons, which will be required in the next section when considering auxiliary spin chain. The $A^*_{2,2+2}(1^+,2^+,3^*,4^*)$ amplitude is given\footnote{See the discussion of $k=2$ case above.} by leg relabeling in equation (\ref{A*(1*,2*,3+,4+)}):
\begin{eqnarray}
A^*_{2,2+2}(1^+,2^+,3^*,4^*) = \frac{1}{\kappa_3^*\kappa_4^*}
\frac{\la p_3 p_4\ra^3}{\la p_4 1\ra\la 1 2\ra\la 2 p_3\ra}\delta^4 (p_1 + p_2 + k_3 + k_4). \label{A*(1+,2+,3*,4*)}
\end{eqnarray}
In the case of $A^*_{3,2+2}(\Omega_1,\Omega_2,g^*_3,g^*_4)$ amplitude the results of the general $k=3$ case considered above give:
\begin{eqnarray}
A^*_{3,2+2}(\Omega_1,\Omega_2,g^*_3,g^*_4) =
\prod_{i=3,5}\frac{\partial^4}{\partial\vleuu^4_i}\Big(~\Omega_{m+4}^2~\Big\{
c_{35} [12345] + c_{36} [12356] + c_{46} [13456]
\Big\}\Big) , \label{A*(1,2,3*,4*)} \nonumber \\
\end{eqnarray}
where
\begin{eqnarray}
c_{35} = \frac{1}{1+\frac{\la p_3\xi_3\ra\la 1 2 3 5\ra}{\la p_3 p_4\ra\la 1234\ra}}, \quad
c_{36} = \frac{1}{1+\frac{\la p_4\xi_4\ra}{\la p_4 1\ra}\frac{\la 1 2 3 5\ra}{\la 2 3 5 6\ra}}, \quad
c_{46} = \frac{1}{1+\frac{\la p_3\xi_3\ra}{\la p_3 p_4\ra}\frac{\la 1 3 5 6\ra}{\la 1 3 4 6\ra}}
\frac{1}{1+\frac{\la p_4\xi_4\ra}{\la p_4 1\ra}\frac{\la 1 3 4 5\ra}{\la 3 4 5 6\ra}} . \nonumber \\
\end{eqnarray}
We have checked, that expressed in terms of helicity spinors this expression reproduces the result of BCFW recursion \cite{vanHamerenBCFW1}.

An interesting question is the spurious pole cancellation. As an example, let us consider $m=3$ case, which is an analog of $n=7$ $k=3$ on-shell amplitude. In the case of the on-shell amplitude we have a term $[12345]$ containing $\la 3451 \ra$ and $\la 5123 \ra$ spurious poles canceling in the sum with the terms $[13456]$ and $[12356]$. In the case of off-shell amplitude the coefficient in front of $[12345]$ term is zero $c_{35}=0$. So, the natural question arises: will this ruin the spurious pole cancellation? Surprisingly, no. In the off-shell case the terms $c_{46}[13456]$ and $c_{36}[12356]$ are regular in the limits $\la 3451 \ra \rightarrow 0$ or $\la 5123 \ra \rightarrow 0$ as
\begin{eqnarray}
c_{46}&=&\frac{1}{1+\frac{\la p_{4}\xi_{4}\ra}{\la p_{4} p_{5}\ra}
	\frac{\la 1346 \ra}{\la 1345 \ra}},\\
c_{36}&=&\frac{1}{1+\frac{\la p_{4}\xi_{4}\ra}{\la p_{4} p_{5}\ra}\frac{\la 1236 \ra}{\la 1235 \ra}}.
\end{eqnarray}

However, unfortunately, more careful analysis indicates that initial conjecture about the  $\Gamma_{tree}$, which is $[1,2\ra$ BCFW contour for on-shell amplitudes is, strictly speaking, incorrect, which can be seen via direct comparison with BCFW recursion results for $A^*_{3,m+2}$ ((\ref{2OffShellgluonsNMHVamplitude}) is valid only for some helicity configurations). Correct contour $\Gamma_{tree}^{off-shell}$ should also include poles associated with $Reg.$ factors.
Taking into account contributions from such poles one can obtain:
\begin{eqnarray}\label{2OffShellgluonsNMHVamplitude1}
A^*_{3,m+2}(\Omega_1,\ldots,\Omega_m,g_{m+1}^*,g_{m+2}^*) &=&
\prod_{i=m+1,m+3}\frac{\partial^4}{\partial\vleuu^4_i}\Big(~\Omega_{m+4}^2~
\omega^3_{m+4}[\Gamma_{tree}^{off-shell}]~\Big) \nonumber \\ &=&
\prod_{i=m+1,m+3}\frac{\partial^4}{\partial\vleuu^4_i}\left(\Omega_{m+4}^2\left\{
\sum_{i <j}c_{ij}[1 i-1ij-1j]
\right\}\right). \nonumber \\
\end{eqnarray}
Here for $c_{ij}$ coefficients we get
\begin{eqnarray}
c_{i,m+3}&=&\frac{1}{1+\frac{\la p_{m+1}\xi_{m+1}\ra}{\la p_{m+1} p_{m+2}\ra}
	\frac{\la 1i-1im+3 \ra}{\la 1i-1im+2 \ra}},\\
c_{i,m+4}&=&\frac{1}{1+\frac{\la p_{m+2}\xi_{m+2}\ra}{\la p_{m+2} 1\ra}\frac{\la 1i-1im+3 \ra}{\la i-1im+3m+4 \ra}},\\
c_{m+2,m+4}&=&\frac{1}{1+\frac{\la p_{m+1}\xi_{m+1}\ra}{\la p_{m+1} p_{m+2}\ra}
	\frac{\la 1 m+1 m+3 m+4\ra}{\la 1 m+1 m+2 m+4 \ra}}
\frac{1}{1+\frac{\la p_{m+2}\xi_{m+2}\ra}{\la p_{m+2} 1\ra}\frac{\la 1 m+1 m+2 m+3 \ra}{\la m+1 m+2 m+3 m+4 \ra}},
\end{eqnarray}
and in other cases $c_{ij}=\hat{A}$, where $\hat{A}$ is shift operator, which shifts supertwistors
$\mathcal{Z}_{m+2},\mathcal{Z}_{m+4}$ as
\begin{eqnarray}
\mathcal{Z}_{m+2}&\mapsto&\mathcal{Z}_{m+2}+\frac{\la p_{m+1}\xi_{m+1}\ra}{\la p_{m+1} p_{m+2}\ra}\mathcal{Z}_{m+3},\\
\mathcal{Z}_{m+4}&\mapsto&\mathcal{Z}_{m+4}+\frac{\la p_{m+2}\xi_{m+2}\ra}{\la p_{m+2} 1\ra}\mathcal{Z}_{1}.
\end{eqnarray}
Such choice of the contour is also consistent with spurious pole cancellation condition.
%\begin{eqnarray}
%c_{i,j}&=&,\\
%c_{i,m+4}&=&\frac{1}{1+\frac{\la p_{m+2}\xi_{m+2}\ra}{\la p_{m+2} 1\ra}\frac{\la 1i-1im+3 \ra}{\la i-1im+3m+4 \ra}},\\
%c_{m+2,m+4}&=&\frac{1}{1+\frac{\la p_{m+1}\xi_{m+1}\ra}{\la p_{m+1} p_{m+2}\ra}
%	\frac{\la 1 m+1 m+3 m+4\ra}{\la 1 m+1 m+2 m+4 \ra}}
%\frac{1}{1+\frac{\la p_{m+2}\xi_{m+2}\ra}{\la p_{m+2} 1\ra}\frac{\la 1 m+1 m+2 m+3 \ra}{\la m+1 m+2 m+3 m+4 \ra}}.
%\end{eqnarray}
Still, the question of the correct choice of integration contour $\Gamma_{tree}$ for the general $m+2n$, $k$ case is open and interesting question.
%The general statement that (\ref{2OffShellgluonsNMHVamplitude}) is free of spurious poles is not (very) easy to formulate, but the discussed examples together with our results for amplitudes with one gluon off-shell make us believe in the self-consistency of the
%presented conjecture.

For $A^*_{4,2+2}(1^-,2^-,g^*_3,g^*_4)$ amplitude $k=4$ and gauge fixed $D$ matrix takes the form
\begin{eqnarray}
D=\left( \begin{array}{cccccc}
1 & 0 & d_{13} & d_{14} & d_{15} & d_{16} \\
0 & 1 & d_{23} & d_{24} & d_{25} & d_{26} \end{array} \right).
\end{eqnarray}
The delta functions constraints completely fix its entries in this case and we get ($i=1,2$):
\begin{eqnarray}
d_{i,3} = - \frac{\la i 4 5 6\ra}{\la 3 4 5 6\ra}, \quad
d_{i,4} =  \frac{\la i 3 5 6\ra}{\la 3 4 5 6\ra}, \quad
d_{i,5} = -\frac{\la i 3 4 6\ra}{\la 3 4 5 6\ra}, \quad
d_{i,6} = \frac{\la i 3 4 5 \ra}{\la 3 4 5 6 \ra} .
\end{eqnarray}
Then the Grassmannian integral for $A^*_{4,2+2}(1^-,2^-,g^*_3,g^*_4)$ evaluates to
\begin{eqnarray}
A^*_{4,2+2}(1^-,2^-,g^*_3,g^*_4) &=& \prod_{i=1,2,3,5}\frac{\partial^4}{\partial\vleuu^4_i}\Bigg(~\Omega_{m+4}^2~
\frac{\la 1 3 4 5\ra\la 1 3 4 6\ra\la 1 3 5 6\ra\la 2 3 4 6\ra\la 2 3 5 6\ra\la 2 4 5 6\ra}{\la 1 2 3 4\ra\la 1 2 3 6\ra\la 1 2 5 6\ra\la 3 4 5 6\ra^3} \nonumber \\
 &\times&
[1 3 4 5 6][2 3 4 5 6]\times \frac{1}{1+\frac{\la p_3\xi_3\ra}{\la p_3 p_4\ra}\frac{\la 1 2 3 5\ra}{\la 1 2 3 4\ra}}\frac{1}{1+\frac{\la p_4\xi_4\ra}{\la p_4 1\ra}\frac{\la 3 4 5 1 \ra}{\la 3 4 5 6\ra}} \Bigg) . \label{A*(1-,2-,3*,4*)}
\end{eqnarray}
We have checked that this expression is consistent with (\ref{A*(1*,2*,3-,4-)}) up to spinor and momenta relabeling.
%In all examples discussed above the choice
%$\Gamma_{tree}=\Gamma_{n+2m}^{[1,2\ra}$ gives correct results, but it will be really surprising that such a choice will indeed work for example for obtaining $A^*_{n,0+2n}$ amplitude from $\Omega^n_{0+2n}$.

\section{Off-shell amplitudes and auxiliary $\mathfrak{gl}(4|4)$ spin chain}

In \cite{offshell-1leg} we have shown how (deformed) gauge invariant amplitudes with one leg off-shell could be described using quantum inverse scattering method (QISM) and auxiliary $\mathfrak{gl}(4|4)$ spin chain. The purpose of this section is to show how this description extends to the case of amplitudes with
multiple off-shell gluons.

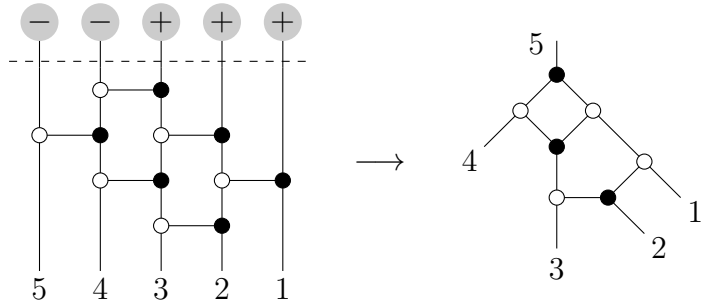
\begin{figure}[htbp]
	\begin{equation*}
	\begin{aligned}
	\begin{tikzpicture}[scale=0.8]
	\drawvline{1}{4.5}
	\drawvline{2}{4.5}
	\drawvline{3}{4.5}
	\drawvline{4}{4.5}
	\drawvline{5}{4.5}
	\drawvacm{1}
	\drawvacm{2}
	\drawvacp{3}
	\drawvacp{4}
	\drawvacp{5}
	\drawbridge{2}{1.5}
	\drawbridge{1}{2.5}
	\drawbridge{3}{2.5}
	\drawbridge{2}{3.5}
	\drawbridge{4}{3.5}
	\drawbridge{3}{4.5}
	\node at (0,-\vacuumheight-4.5*\bridgedistance-\labelvdist) {5};
	\node at (1,-\vacuumheight-4.5*\bridgedistance-\labelvdist) {4};
	\node at (2,-\vacuumheight-4.5*\bridgedistance-\labelvdist) {3};
	\node at (3,-\vacuumheight-4.5*\bridgedistance-\labelvdist) {2};
	\node at (4,-\vacuumheight-4.5*\bridgedistance-\labelvdist) {1};
	\draw[dashed] (-0.5,-1.2*\bridgedistance) -- (4.5,-1.2*\bridgedistance) ;
	\end{tikzpicture}
	\end{aligned}
	\quad\longrightarrow\quad
	\begin{aligned}
	\begin{tikzpicture}[scale=0.9]
	\draw (0,-0) -- (0,0.5); % Minimal form factor
	\draw (0,0) -- (-\hdist,-\hdist) -- (0,-2*\hdist) -- (+\hdist,-\hdist) -- (0,0); % Box
	\draw (0,-2*\hdist) -- (0,-2*\hdist-\ddist) -- (\ddist,-2*\hdist-\ddist) -- (\hdist+\ddist,-\hdist-\ddist) -- (+\hdist,-\hdist); % Pentagon
	\draw (+\hdist+\ddist,-\hdist-\ddist) -- (+2*\hdist+\ddist,-2*\hdist-\ddist);
	\draw (-\hdist,-\hdist) -- (-2*\hdist,-2*\hdist);
	\draw (0,-2*\hdist-\ddist) -- (0,-2*\hdist-2*\ddist);
	\draw (\ddist,-2*\hdist-\ddist) -- (+\hdist+\ddist,-3*\hdist-\ddist);
	\node[db] at (0,-0) {};
	\node[dw] at (+\hdist,-\hdist) {};
	\node[db] at (0,-2*\hdist) {};
	\node[dw] at (-\hdist,-\hdist) {};
	\node[dw] at (0,-2*\hdist-\ddist) {};
	\node[db] at (\ddist,-2*\hdist-\ddist) {};
	\node[dw] at (\hdist+\ddist,-\hdist-\ddist) {};
	\node at (+2*\hdist +\ddist +\labelddist,-2*\hdist-\ddist-\labelddist) {1};
	\node at (-\labelhdist,1.5*\labelvdist) {$5$};
	\node at (-2*\hdist-\labelddist,-2*\hdist-\labelddist) {4};
	\node at (0,-2*\hdist-2*\ddist-\labelvdist) {3};
	\node at (\hdist +\ddist +\labelddist,-3*\hdist-\ddist-\labelddist) {2};
	\end{tikzpicture}
	\end{aligned}
	\end{equation*}
	\caption{On-shell diagram construction via BCFW bridges for $A_{5}$ amplitude
	}
	\label{fig: A5}
\end{figure}

Originally auxiliary spin chain description of the on-shell tree-level amplitudes appeared as a result of investigations of their symmetry properties. First, the Yangian symmetry, combining  invariance under superconformal and dual superconformal transformations \cite{DualSuperConformalSymmetry} was proven for on-shell tree-level amplitudes in \cite{YangianSymmetryTreeAmplitudes}. Next, it was claimed \cite{Drummond_Grassmannians_Tduality,Drummond_Yangian_origin_Grassmannian_integral} that the Grassmannian integral representation for on-shell amplitudes (\ref{GrassmannianOnshell}) is the most general form of rational Yangian invariant. The study of tree-level scattering amplitudes within the context of QISM was started in \cite{AmplitudesSpectralParameter1,AmplitudesSpectralParameter2}, with the introduction of the notion of spectral parameter, which was later interpreted as a deformed particle helicity. Next, the authors of \cite{Frassek_BetheAnsatzYangianInvariants,Chicherin_YangBaxterScatteringAmplitudes} proposed to study certain auxiliary spin chain monodromies. The introduced monodromies depended on an extra auxiliary spectral parameter, while the spectral parameters of \cite{AmplitudesSpectralParameter1,AmplitudesSpectralParameter2} played the role of spin chain inhomogeneities.  Yangian invariants and thus on-shell amplitudes are then found as the eigenstates of these monodromies. Further,  \cite{Kanning_ShortcutAmplitudesIntegrability,Broedel_DictionaryRoperatorsOnshellGraphsYangianAlgebras} provided a systematic classification of Yangian invariants obtained within QISM. Later QISM description was extended to form factors\footnote{See also \cite{SoftTheoremsFormFactors} for preliminary steps.} \cite{FormFactorsGrassmanians} and amplitudes with one leg off-shell \cite{offshell-1leg}. It should be noted, that in the last two cases the Yangian invariance is explicitly broken by the corresponding vacuum states. Still, the machinery of QISM could be applied in those cases also.

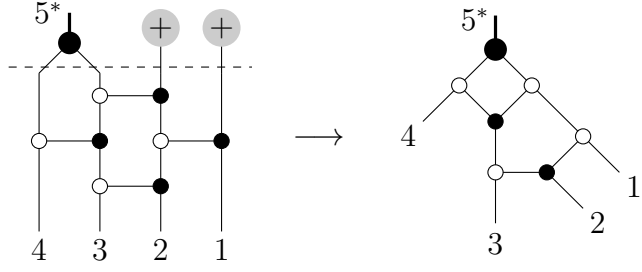
\begin{figure}[htbp]
	\begin{equation*}
	\begin{aligned}
	\begin{tikzpicture}[scale=0.8]
	\drawvline{1}{3.5}
	\drawvline{2}{3.5}
	\drawvline{3}{3.5}
	\drawvline{4}{3.5}
	\drawminimalA{1}
	\drawvacp{3}
	\drawvacp{4}
	\drawbridge{2}{1.5}
	\drawbridge{1}{2.5}
	\drawbridge{3}{2.5}
	\drawbridge{2}{3.5}
	\node[offshell] at (0.5,-0.5) {};
	\node at (\labelvdist/2,0) {$5^*$};
	\node at (0,-\vacuumheight-3.5*\bridgedistance-\labelvdist) {4};
	\node at (1,-\vacuumheight-3.5*\bridgedistance-\labelvdist) {3};
	\node at (2,-\vacuumheight-3.5*\bridgedistance-\labelvdist) {2};
	\node at (3,-\vacuumheight-3.5*\bridgedistance-\labelvdist) {1};
	\draw[dashed] (-0.5,-1.2*\bridgedistance) -- (3.5,-1.2*\bridgedistance) ;
	\end{tikzpicture}
	\end{aligned}
	\quad\longrightarrow\quad
	\begin{aligned}
	\begin{tikzpicture}[scale=0.9]
	\draw[very thick] (0,-0) -- (0,0.5); % Minimal form factor
	\draw (0,0) -- (-\hdist,-\hdist) -- (0,-2*\hdist) -- (+\hdist,-\hdist) -- (0,0); % Box
	\draw (0,-2*\hdist) -- (0,-2*\hdist-\ddist) -- (\ddist,-2*\hdist-\ddist) -- (\hdist+\ddist,-\hdist-\ddist) -- (+\hdist,-\hdist); % Pentagon
	\draw (+\hdist+\ddist,-\hdist-\ddist) -- (+2*\hdist+\ddist,-2*\hdist-\ddist);
	\draw (-\hdist,-\hdist) -- (-2*\hdist,-2*\hdist);
	\draw (0,-2*\hdist-\ddist) -- (0,-2*\hdist-2*\ddist);
	\draw (\ddist,-2*\hdist-\ddist) -- (+\hdist+\ddist,-3*\hdist-\ddist);
	\node[offshell] at (0,-0) {};
	\node[dw] at (+\hdist,-\hdist) {};
	\node[db] at (0,-2*\hdist) {};
	\node[dw] at (-\hdist,-\hdist) {};
	\node[dw] at (0,-2*\hdist-\ddist) {};
	\node[db] at (\ddist,-2*\hdist-\ddist) {};
	\node[dw] at (\hdist+\ddist,-\hdist-\ddist) {};
	\node at (+2*\hdist +\ddist +\labelddist,-2*\hdist-\ddist-\labelddist) {1};
	\node at (-\labelhdist,1.5*\labelvdist) {$5^*$};
	\node at (-2*\hdist-\labelddist,-2*\hdist-\labelddist) {4};
	\node at (0,-2*\hdist-2*\ddist-\labelvdist) {3};
	\node at (\hdist +\ddist +\labelddist,-3*\hdist-\ddist-\labelddist) {2};
	\end{tikzpicture}
	\end{aligned}
	\end{equation*}
	\caption{On-shell diagram construction via BCFW bridges for $A^*_{4+1}$ amplitude
	}
	\label{fig: A4+1}
\end{figure}

The auxiliary $\mathfrak{gl}(4|4)$ spin chain in the case of on-shell amplitudes arises by writing Yangian invariance condition as a system of eigenvalue equations for the elements of a suitable monodromy matrix $M(u)$ \cite{Frassek_BetheAnsatzYangianInvariants,Chicherin_YangBaxterScatteringAmplitudes,Kanning_ShortcutAmplitudesIntegrability}:
\begin{eqnarray}
M_{ab}(u)|\Psi\rangle = {C}_{ab}|\Psi\rangle. \label{monodromyEq}
\end{eqnarray}
Here $u$ is the auxiliary spectral parameter and $C_{ab}$ are monodromy eigenvalues.  The monodromy eigenvectors $|\Psi\rangle$ are the elements of the Hilbert space $V = V_1\otimes\ldots\otimes V_n$ with $V_i$ being a particular $\mathfrak{gl}(4|4)$ non-compact representation built using a single family of Jordan-Schwinger harmonic superoscillators $\overline{\bf w}^\mathcal{A}, {\bf w}^\mathcal{B}$, $\mathcal{A},\mathcal{B} = 1\ldots 8$. The latter could be conveniently written in terms of Heisenberg pairs
\begin{eqnarray}
J^{\mathcal{AB}} = \overline{\bf w}^\mathcal{A} {\bf w}^\mathcal{B} = x^\mathcal{A} p^\mathcal{B},\quad x^\mathcal{A} = \left(\vll^{\alpha},-\frac{\partial}{\partial\vlt^{\dotalpha}}, \frac{\partial}{\partial\vlet^A} \right) , \quad p^\mathcal{A} = \left(\frac{\partial}{\partial\vll^{\alpha}}, \vlt^{\dotalpha} , \vlet^A\right),
\end{eqnarray}
with $[x^{\mathcal{A}}, p^{\mathcal{B}}\} = (-1)^{|\mathcal{A}|}\delta^{\mathcal{AB}}$. Here $[\cdot , \cdot\}$ denotes graded commutator and $|\cdot|$ - grading. A vacuum state required in the construction of Yangian invariants $|\Psi\rangle_{n,k}$ corresponding to the on-shell N$^{k-2}$MHV $n$-point tree-level amplitudes $A_{n,k}$ is given by
\begin{eqnarray}
|{\bf 0}\rangle_{k,n} = \delta_1^+ \cdots \delta_{n-k}^+\delta_{n-k+1}^-\cdots \delta_n^- ,
\end{eqnarray}
where $\delta_i^+ \equiv \delta^2 (\vll_i)$ is the vacuum for the positive helicity state at position $i$ and $\delta_i^- \equiv \delta^2 (\vlt_i)\delta^4 (\vlet_i)$ is the corresponding vacuum for negative helicity state. In the following we will also need a graphical notation for the above vacuum states introduced in \cite{FormFactorsGrassmanians}:
\begin{equation}
\begin{aligned}\begin{tikzpicture}[scale=0.8]
\drawvacp{1}
\node[dl] at (0,-\vacuumheight-0*\bridgedistance-\labelvdist) {$i$};
\end{tikzpicture}\end{aligned}
=
\delta^+_{i}=\delta^{2}(\vll_i)\eqncom\qquad
\begin{aligned}\begin{tikzpicture}[scale=0.8]
\drawvacm{1}
\node[dl] at (0,-\vacuumheight-0*\bridgedistance-\labelvdist) {$i$};
\end{tikzpicture}\end{aligned}
=
\delta^-_{i}=\delta^{2}(\vlt_i)\delta^{4}(\vlet_i)
\eqndot
\label{fig: vacua}
\end{equation}
The monodromy matrix of the auxiliary spin chain expressed in terms of Lax operators reads
\begin{eqnarray}
M(u,\{v_i\}) = \mathcal{L}_1 (u,v_1)\ldots\mathcal{L}_k (u,v_k)
\mathcal{L}_{k+1} (u,v_{k+1})\ldots\mathcal{L}_n (u,v_n), \label{monodromy-matrix}
\end{eqnarray}
where $v_i$ are spin chain inhomogeneities and Lax operators $\mathcal{L}_i (u,v)$ are given by
\begin{eqnarray}
\mathcal{L} (u,v) = u - v + \sum_{a,b} e_{ab} J_{ba}\, . \label{LaxOperator}
\end{eqnarray}
Here, the matrix $e_{ab}$ acting in the auxiliary space is given by $(e_{ab})_{cd} = \delta_{ac}\delta_{bd}$ and the action of Lax operators on vacuum states is given by
\begin{eqnarray}
\mathcal{L}_i (u)\;\delta_i^+ = (u-1)\; \mathbb{I}\;\delta_i^+, \qquad
\mathcal{L}_i (u)\;\delta_i^- = u\;\mathbb{I}\;\delta_i^- . \label{Lax-vacua}
\end{eqnarray}
The solution of the eigenvalue equation (\ref{monodromyEq}) provides us with the expressions for Yangian invariants labeled by the permutations $\sigma$ with minimal\footnote{It means, that there is no other decomposition of $\sigma$ into a smaller number of transpositions.} decomposition $\sigma = (i_1,j_1)\ldots (i_P,j_P)$ \cite{Chicherin_YangBaxterScatteringAmplitudes,Kanning_ShortcutAmplitudesIntegrability,Broedel_DictionaryRoperatorsOnshellGraphsYangianAlgebras}:
\begin{eqnarray}
|\Psi\rangle = \mathcal{R}_{i_1,j_1}(\bar{u}_1)\ldots\mathcal{R}_{i_Pj_P}(\bar u_P)|{\bf 0}\rangle_{k,n}
\end{eqnarray}
where \cite{Chicherin_YangBaxterScatteringAmplitudes} (see also \cite{Frassek_BetheAnsatzYangianInvariants})
\begin{eqnarray}
\mathcal{R}_{ij} (u) = \Gamma (-u)(x_j\cdot p_i)^u = \int_0^\infty \frac{d\alpha}{\alpha^{1+u}}e^{-\alpha (x_j\cdot p_i)} . \label{RoperatorDef}
\end{eqnarray}
Here $\Gamma$ is the Euler gamma function and
\begin{eqnarray}
\bar u_p = v_{\tau_p (i_p)} - v_{\tau_p  (j_p)},\qquad  \tau_p=\tau_{p-1}\circ(i_p,j_p)=(i_1,j_1)\cdots(i_p,j_p).
\end{eqnarray}

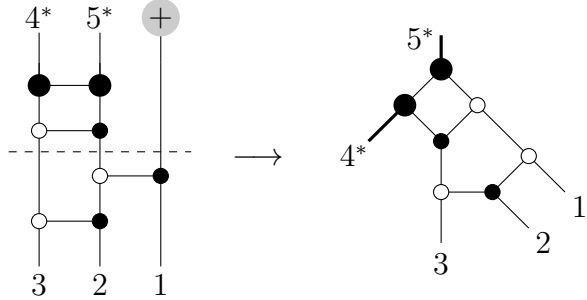
\begin{figure}[htbp]
	\begin{equation*}
		\begin{aligned}
			\begin{tikzpicture}[scale=0.8]
			\drawvline{1}{4.5}
			\drawvline{2}{4.5}
			\drawvline{3}{4.5}
			\drawvacp{3}
			\drawnumline{2}{$5^*$}
			\drawnumline{1}{$4^*$}
			\drawoffshellbridge{1}{1.5}
			\drawbridge{1}{2.5}
			\drawbridge{2}{3.5}
			\drawbridge{1}{4.5}
			\node at (0,-\vacuumheight-4.5*\bridgedistance-\labelvdist) {3};
			\node at (1,-\vacuumheight-4.5*\bridgedistance-\labelvdist) {2};
			\node at (2,-\vacuumheight-4.5*\bridgedistance-\labelvdist) {1};
			\draw[dashed] (-0.5,-3.3*\bridgedistance) -- (2.5,-3.3*\bridgedistance) ;
			\end{tikzpicture}
		\end{aligned}
		\quad\longrightarrow\quad
		\begin{aligned}
			\begin{tikzpicture}[scale=0.9]
			\draw[very thick] (0,-0) -- (0,0.5); % Minimal form factor
			\draw (0,0) -- (-\hdist,-\hdist) -- (0,-2*\hdist) -- (+\hdist,-\hdist) -- (0,0); % Box
			\draw (0,-2*\hdist) -- (0,-2*\hdist-\ddist) -- (\ddist,-2*\hdist-\ddist) -- (\hdist+\ddist,-\hdist-\ddist) -- (+\hdist,-\hdist); % Pentagon
			\draw (+\hdist+\ddist,-\hdist-\ddist) -- (+2*\hdist+\ddist,-2*\hdist-\ddist);
			\draw[very thick] (-\hdist,-\hdist) -- (-2*\hdist,-2*\hdist);
			\draw (0,-2*\hdist-\ddist) -- (0,-2*\hdist-2*\ddist);
			\draw (\ddist,-2*\hdist-\ddist) -- (+\hdist+\ddist,-3*\hdist-\ddist);
			\node[offshell] at (0,-0) {};
			\node[dw] at (+\hdist,-\hdist) {};
			\node[db] at (0,-2*\hdist) {};
			\node[offshell] at (-\hdist,-\hdist) {};
			\node[dw] at (0,-2*\hdist-\ddist) {};
			\node[db] at (\ddist,-2*\hdist-\ddist) {};
			\node[dw] at (\hdist+\ddist,-\hdist-\ddist) {};
			\node at (+2*\hdist +\ddist +\labelddist,-2*\hdist-\ddist-\labelddist) {1};
			\node at (-\labelhdist,1.5*\labelvdist) {$5^*$};
			\node at (-2*\hdist-\labelddist,-2*\hdist-\labelddist) {$4^*$};
			\node at (0,-2*\hdist-2*\ddist-\labelvdist) {3};
			\node at (\hdist +\ddist +\labelddist,-3*\hdist-\ddist-\labelddist) {2};
			\end{tikzpicture}
		\end{aligned}
	\end{equation*}
	\caption{On-shell diagram construction via BCFW bridges for $A^{*}_{3+2}$ amplitude
	}
	\label{fig: A3+2}
\end{figure}

To describe amplitudes with one leg off-shell in \cite{offshell-1leg} we required one additional ingredient - the vacuum state corresponding to minimal off-shell amplitude:
\begin{eqnarray}
A^*_{2,2+1}(2,3) = A^*_{2,2+1}(g^*_1,2,3) = \frac{1}{\kappa^*}\frac{\la 2 3\ra}{\la p 2\ra\la p 3\ra}
\delta^2 (\vltu_2)\delta^2 (\vltu_3)\delta^4 (\vleu_2)\delta^4 (\vleu_3) ,
\end{eqnarray}
where ($k$ is the off-shell gluon momentum and $p$ is its direction)
\begin{eqnarray}
\vltu_2 = \vlt_2 + \frac{\la 3|k}{\la 3 2\ra},
\qquad \vltu_3 = \vlt_3 + \frac{\la 2|k}{\la 2 3\ra}
\qquad \vleu_2 = \vlet_2 + \frac{\la p 3\ra}{\la 2 3\ra}\vlet_p ,  \qquad \vleu_3 = \vlet_3 + \frac{\la p 2\ra}{\la 3 2\ra}\vlet_p . \nonumber \\ \label{lambda-eta-offshell-vertex}
\end{eqnarray}
Then, for example the deformed\footnote{It is not clear what is the meaning of the deformation in the off-shell case as the Yangian invariance is broken now, while the deformation was originally introduced to obtain the general Yangian invariant expressions.} off-shell amplitude $A^*_{2,3+1}$ could be written as \cite{offshell-1leg}:
\begin{eqnarray}
A^*_{2,3+1}(\bar u_1,\bar u_2) = \mathcal{R}_{23} (\bar u_1)\mathcal{R}_{12} (\bar u_2)
\delta^2 (\vll_1)\frac{1}{\kappa^*}\frac{\la 2 3\ra}{\la p 2\ra\la p  3\ra}\delta^2 (\vltu_2)\delta^2 (\vltu_3)\delta^4 (\vleu_2) \delta^4 (\vleu_3) ,
\end{eqnarray}
where $\vllu_i$, $\vleu_i$ are defined in (\ref{lambda-eta-offshell-vertex}) and $\bar u_1 = v_{32} = v_3 - v_2$, $\bar u_2 = v_{31} = v_3 - v_1$. Using definition of $\mathcal{R}$ operators (\ref{RoperatorDef}) we get \cite{offshell-1leg}:
\begin{multline}
A_{2,3+1}^*(v_1,v_2,v_3) =
\frac{1}{\kappa^{*}}\int\frac{d\alpha_2}{\alpha_2}
\int\frac{d\alpha_1}{\alpha_1}\frac{\la 2 3\ra}{\la p 2\ra\la p 3\ra \left(1-\alpha_2\frac{\la p 3\ra}{\la p 2\ra}\right)}\times  \\  \times \delta^2 (\vll_1 - \alpha_1\vll_2 + \alpha_1\alpha_2\vll_3)
\delta^2 (\vltu_2 + \alpha_1\vlt_1) \delta^4 (\vleu_2 + \alpha_1 \vlet_1)
\delta^2 (\vltu_3 + \alpha_2\vltu_2)\delta^4 (\vleu_3 + \alpha_2\vleu_2)  \\
=  \frac{1}{\kappa^*}\frac{1}{\la p 1\ra\la 1 2\ra\la 2 3\ra\la 3 p\ra}\left(\frac{\la 2 3\ra}{\la 1 3\ra}\right)^{v_{31}}
\left(\frac{\la 1 3\ra}{\la 1 2\ra}\right)^{v_{32}}
\delta^4 (\sum_{i=1}^3\vll_i\vlt_i + k)
\delta^8 (\sum_{i=1}^3\vll_i\vlet_i + \vll_p\vlet_p) . \\
\end{multline}
The off-shell amplitude $A^*_{2,3+1}$ is recovered by setting  deformation parameters to zero $v_i = 0$.

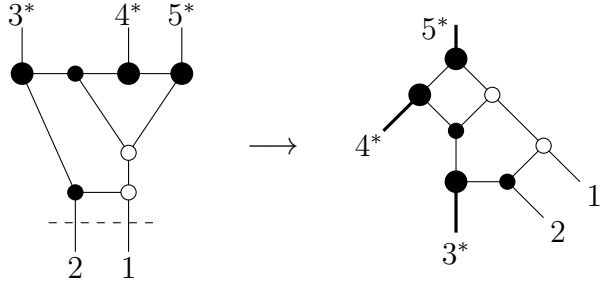
\begin{figure}[htbp]
	\begin{equation*}
	\begin{aligned}
	\begin{tikzpicture}[scale=0.7]
	\drawnumline{4}{$5^*$}
	\drawnumline{3}{$4^*$}
	\drawnumline{1}{$3^*$}
	\node[offshell] at (3,-\vacuumheight -0.5*\bridgedistance) {};
	\node[offshell] at (2,-\vacuumheight -0.5*\bridgedistance) {};
	\node[db] at (1,-\vacuumheight -0.5*\bridgedistance) {};
    \node[offshell] at (0,-\vacuumheight -0.5*\bridgedistance) {};

	\draw (0,-\vacuumheight-0.5*\bridgedistance) -- (3,-\vacuumheight-0.5*\bridgedistance);
	\draw (1,-\vacuumheight-0.5*\bridgedistance) --
	(2,-\vacuumheight -2.5*\bridgedistance);
	\draw (3,-\vacuumheight-0.5*\bridgedistance) --
	(2,-\vacuumheight -2.5*\bridgedistance);
	\draw (2,-\vacuumheight -2.5*\bridgedistance) --
	(2,-\vacuumheight -3.5*\bridgedistance);
	\draw (0,-\vacuumheight -0.5*\bridgedistance) --
	(1,-\vacuumheight -3.5*\bridgedistance);
	\draw (1,-\vacuumheight -3.5*\bridgedistance) --
	(2,-\vacuumheight -3.5*\bridgedistance);
	\draw (1,-\vacuumheight -3.5*\bridgedistance) --
	      (1,-\vacuumheight -5*\bridgedistance);
	\draw (2,-\vacuumheight -3.5*\bridgedistance) --
	      (2,-\vacuumheight -5*\bridgedistance);
	
	\node[dw] at (2,-\vacuumheight -2.5*\bridgedistance) {};
	\node[dw] at (2,-\vacuumheight -3.5*\bridgedistance) {};
	\node[db] at (1,-\vacuumheight -3.5*\bridgedistance) {};
	\node at (1,-\vacuumheight-5*\bridgedistance-\labelvdist) {2};
	\node at (2,-\vacuumheight-5*\bridgedistance-\labelvdist) {1};
	
	\draw[dashed] (0.5,-5.6*\bridgedistance) -- (2.5,-5.6*\bridgedistance) ;
	\end{tikzpicture}
	\end{aligned}
	\quad\longrightarrow\quad
	\begin{aligned}
	\begin{tikzpicture}[scale=0.9]
	\draw[very thick] (0,-0) -- (0,0.5); % Minimal form factor
	\draw (0,0) -- (-\hdist,-\hdist) -- (0,-2*\hdist) -- (+\hdist,-\hdist) -- (0,0); % Box
	\draw (0,-2*\hdist) -- (0,-2*\hdist-\ddist) -- (\ddist,-2*\hdist-\ddist) -- (\hdist+\ddist,-\hdist-\ddist) -- (+\hdist,-\hdist); % Pentagon
	\draw (+\hdist+\ddist,-\hdist-\ddist) -- (+2*\hdist+\ddist,-2*\hdist-\ddist);
	\draw[very thick] (-\hdist,-\hdist) -- (-2*\hdist,-2*\hdist);
	\draw[very thick] (0,-2*\hdist-\ddist) -- (0,-2*\hdist-2*\ddist);
	\draw (\ddist,-2*\hdist-\ddist) -- (+\hdist+\ddist,-3*\hdist-\ddist);
	\node[offshell] at (0,-0) {};
	\node[dw] at (+\hdist,-\hdist) {};
	\node[db] at (0,-2*\hdist) {};
	\node[offshell] at (-\hdist,-\hdist) {};
	\node[offshell] at (0,-2*\hdist-\ddist) {};
	\node[db] at (\ddist,-2*\hdist-\ddist) {};
	\node[dw] at (\hdist+\ddist,-\hdist-\ddist) {};
	\node at (+2*\hdist +\ddist +\labelddist,-2*\hdist-\ddist-\labelddist) {1};
	\node at (-\labelhdist,1.5*\labelvdist) {$5^*$};
	\node at (-2*\hdist-\labelddist,-2*\hdist-\labelddist) {$4^*$};
	\node at (0,-2*\hdist-2*\ddist-\labelvdist) {$3^*$};
	\node at (\hdist +\ddist +\labelddist,-3*\hdist-\ddist-\labelddist) {2};
	\end{tikzpicture}
	\end{aligned}
	\end{equation*}
	\caption{On-shell diagram construction via BCFW bridges for $A^{*}_{2+3}$ amplitude
	}
	\label{fig: A2+3}
\end{figure}

To see what should be done to apply QISM machinery to amplitudes with multiple off-shell gluons let us recall that we are actually have a systematic procedure to construct a given on-shell diagram starting from its corresponding permutation \cite{AmplitudesPositiveGrassmannian}. The latter procedure is known as a BCFW bridge addition construction. First, the permutation is decomposed into a chain of consequent transpositions. Then each transposition $(i,j)$ is interpreted as a BCFW bridge. Finally, the obtained BCFW bridges are applied to a corresponding empty vacuum diagram with the prescribed values of $k$ and $n$\footnote{See \cite{AmplitudesPositiveGrassmannian} for more details.}. The BCFW bridge addition operation is given by
\begin{gather}\label{bridgeadditionoperation}
 \begin{tikzpicture}[scale=0.8,baseline=-43pt]
 \drawvline{1}{2}
 \drawvline{2}{2}
 \drawbridge{1}{2}
 \node[dl] at (0,-\vacuumheight-2*\bridgedistance-\labelvdist) {$j$};
 \node[dl] at (1,-\vacuumheight-2*\bridgedistance-\labelvdist) {$i$};
 \end{tikzpicture} f(\vll_i , \vlt_i , \vlet_i , \vll_j , \vlt_j , \vlet_j  ) = R_{i j} f(\vll_i , \vlt_i , \vlet_i , \vll_j , \vlt_j , \vlet_j  )
= \nonumber \\
\int \frac{d\alpha}{\alpha} f (\vll_i - \alpha\vll_j , \vlt_i, \vlet_i,
\vll_j , \vlt_j + \alpha\vlt_i , \vlet_j + \alpha\vlet_i)
\end{gather}
It is precisely the steps we are going through within QISM approach, the only difference is that BCFW bridges or $\mathcal{R}$ operators are deformed now. So, to get QISM description of a given on-shell diagram corresponding to some factorization channel of the off-shell amplitude under consideration we divide it into a vacuum state and a sequence of bridge additions or $\mathcal{R}$ operators acting on it.  The examples of this division are shown in Figs. \ref{fig: A5}-\ref{fig: A2+3}. The large black vertexes in the above figures correspond to minimal off-shell vertexes (see Fig. \ref{fig: offshell-vertex}), while small black and white vertexes to usual 3-point MHV and $\overline{\text{MHV}}$ vertexes correspondingly. The easiest way to get explicit expressions of vacuum states, corresponding to the parts of on-shell diagrams above the dashed line (see Figs. \ref{fig: A5}-\ref{fig: A2+3}), is to either use off-shell BCFW recursion \cite{vanHamerenWL1,vanHamerenBCFW2} or our Grassmannian representation. This way in the case of amplitudes with two off-shell gluons  vacuum states are given (restoring the dependence on Grassmann variables for on-shell states when needed) by the sum of $A^*_{2,2+2}(1^+,2^+,g^*_3,g^*_4)$ (\ref{A*(1+,2+,3*,4*)}), $A^*_{3,2+2}(\Omega_1,\Omega_2,g^*_3,g^*_4)$ (\ref{A*(1,2,3*,4*)}) and $A^*_{4,2+2}(1^-,2^-,g^*_3,g^*_4)$ (\ref{A*(1-,2-,3*,4*)}) multiplied by the required number of vacuum states for extra on-shell states (\ref{fig: vacua}).

Finally, in out previous paper \cite{offshell-1leg} we have shown, that gauge invariant amplitudes with one leg off-shell are no longer eigenvectors of monodromy matrix of the auxiliary $\mathfrak{gl}(4|4)$ spin chain. The latter, however, turn out to be eigenvectors of corresponding transfer matrix. The last property was the consequence of multiplicative renormalizability of amplitudes with one leg off-shell. Thus, in the case of multiple off-shell gluons we are again expecting amplitudes to have the same properties, in particular they should be eigenvectors of transfer matrix.

\section{Conclusion}
In this paper we presented a conjecture for Grassmannian integral representation of $\mathcal{N}=4$ SYM tree level gauge invariant off-shell amplitudes containing arbitrary number of off-shell gluons or equivalently Wilson line form factors with an  arbitrary number of Wilson line operator insertions. The conjecture was successfully verified on multiple examples known in the literature \cite{vanHamerenBCFW1}. We have also derived some new closed formulas for off-shell amplitudes with arbitrary number of on-shell particles and fixed number of off-shell gluons. In addition we discuss the relation of our Grassmannian representation with the integrability approach to the
amplitudes of $\mathcal{N}=4$ SYM.

It is remarkable that within our approach we can obtain Grassmannian integral representation for amplitudes  without on-shell particles et all, i.e. for correlation functions of Wilson line operators (pure off-shell objects). This observation leads us to conjecture that in fact all gauge invariant observables in $\mathcal{N}=4$ SYM (scattering amplitudes, form factors and correlation functions of various gauge invariant operators, not necessary local) can be uniformly
represented in one way or another in terms of integrals over Grassmannian or its subsets.

There are some open questions and possible further developments along the lines considered in this article. First, it would be interesting to consider supersymmetrized version of Wilson line operators. So far we have treated all on-shell states in manifestly supersymmetric way, while the off-shell states
were restricted to gluons only. It is tempting to claim that the fully supersymmetric version of off-shell amplitude, where both on-shell states and Wilson line operators are treated
in supersymmetric way, will be given just by $\Omega_{m+2n}^k[\Gamma]$ without any constraints on the Grassmann counterparts of $2n$ helicity spinor variables
parameterizing $n$ off-shell momenta $k_i$. However, we think that more accurate consideration is necessary.

Next, it would be very interesting to fully uncover geometrical picture behind conjectured here Grassmannian representation for off-shell amplitudes as well as similar representations
for form factors \cite{FormFactorsGrassmanians,q2zeroFormFactors,BrandhuberConnectedPrescription,HeConnectedPrescription}. It is interesting to see if the ``Amplituhedron'' picture can be extended to all possible gauge invariant observables in  $\mathcal{N}=4$ SYM. In addition the combinatorial nature of modifications to on-shell diagram formalism used here as well as in \cite{FormFactorsGrassmanians,q2zeroFormFactors,offshell-1leg} remains mostly unexplored.

Despite the relation between integrable systems and Grassmannian representation of off-shell amplitudes discussed here and in \cite{offshell-1leg} some important questions remain unanswered.
Namely, the present relation allows us to use auxiliary spin chain to add on-shell legs to the amplitudes and it would be interesting if there is a similar approach based on quantum inverse scattering method which would allow us to add additional off-shell legs.

In conclusion, we would like to note that it would be interesting to extended scattering equations and ambitwistor string approaches for the case of gauge invariant off-shell amplitudes considered here. An important topic is the calculation of loop corrections to gauge invariant off-shell amplitudes. Also, it is extremely interesting to see how the ideas presented here and in \cite{offshell-1leg} work in other theories, for example in gravity and supergravity, where we have also a well developed approach based on high-energy effective lagrangian \cite{gravityEL1,gravityEL2,gravityEL3,gravityEL4,gravityEL5}, see also \cite{gravityRegge1,gravityRegge2,gravityRegge3,gravityRegge4,gravityRegge5} for similar research along this direction.

\section*{Acknowledgements}

The authors would like to thank D.I. Kazakov and S.E. Derkachov both for drawing our attention to this problem as well  as for interesting and stimulating discussions. A.O would like to thank L.N. Lipatov for explanations and discussions on the subject of effective high energy lagrangian and L.V. would like to thank Yu-tin Huang for interesting and stimulating discussion. This work was supported by RSF grant \#16-12-10306.
%This work was supported by RFBR grants \# 14-02-000494 , \# 16-02-00943 and contract \# 02.A03.21.0003 from 27.08.2013 with Russian Ministry of Science and Education.
%

\appendix

\section{$A^*_{1+2}(1^+,2^*,3^*)$ and on-shell diagrams}\label{appA}

In this appendix we present the details of the calculation of the on-shell diagram corresponding to $A^*_{1+2}(1^+,2^*,3^*)$ amplitude. This particular example is the check of our hypotheses about the structure of on-shell diagrams for off-shell amplitudes. Namely, we expect that the on-shell diagrams for off-shell amplitudes could be obtained from the corresponding on-shell diagrams for on-shell amplitudes by substituting on-shell vertexes containing off-shell legs with the minimal off-shell vertexes (see Fig. \ref{fig: offshell-vertex}). So, the on-shell
diagrams for off-shell gauge invariant amplitudes should be given by a combination of ordinary $\mbox{MHV}_3$, $\overline{\mbox{MHV}}_3$ on-shell amplitudes together with one additional off-shell vertex $A^*_{1+2}$ which should be glued with other parts of the on-shell diagram via on-shell legs only.

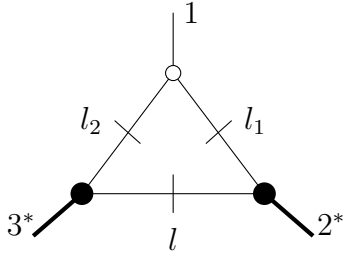
\begin{figure}[htbp]
	\begin{equation*}
	\begin{aligned}
	\begin{tikzpicture}[scale=0.8]
	\draw (0,0) -- (1.5,-2);
	\draw (0,0) -- (-1.5,-2);
	\draw (-1.5,-2) -- (1.5,-2);
	\draw (0,0) -- (0,1);
	\draw[ultra thick] (1.5,-2) -- (2.3,-2.7);
    \draw[ultra thick] (-1.5,-2) -- (-2.3,-2.7);
    \draw (0,-2+1.5*\labelddist) -- (0,-2-1.5*\labelddist);
    \draw (1.5/2+\labelddist,-1+\labelddist) -- (1.5/2-\labelddist,-1-0.8*\labelddist);
    \draw (-1.5/2-\labelddist,-1+\labelddist) -- (-1.5/2+\labelddist,-1-0.8*\labelddist);
	\node[dw] at (0,0) {};
	\node[offshell] at (1.5,-2) {};
	\node[offshell] at (-1.5,-2) {};
	\node at (\labelhdist,1) {1};
	\node at (2.3+\labelhdist,-2.5) {$2^*$};
    \node at (-2.2-\labelhdist,-2.5) {$3^*$};
    \node at (1.5/2+2*\labelhdist,-1+\labelddist) {$l_1$};
    \node at (-1.5/2-2*\labelhdist,-1+\labelddist) {$l_2$};
    \node at (0,-2-3.5*\labelddist) {$l$};
	\end{tikzpicture}
	\end{aligned}
	\end{equation*}
	\caption{On-shell diagram with off-shell vertexes for $A^*_{1+2}(1^+,2^*,3^*)$ amplitude
	}
	\label{fig: A1+2-appendix1}
\end{figure}

Let's see whether this prescription will reproduce us $A^*_{1+2}(1^+,2^*,3^*)$ amplitude. Corresponding
on-shell diagram is shown in Fig. \ref{fig: A1+2-appendix1} and its expression is given by (the action of projectors $\partial^4/\partial^4 \tilde{\eta}_{p_2}$ and $\partial^4/\partial^4 \tilde{\eta}_{p_3}$ is assumed):
\begin{eqnarray}
\Omega=\int \prod_{I=l,l_1,l_2}
\frac{d^2\lambda_I~d^2\tilde{\lambda}_I~d^4\tilde{\eta}_I}{U(1)}
A^*_{1+2}(3^*,l_2,l)A_{3}^{\overline{MHV}}(l_2,1,l_1)A^*_{1+2}(l_1,2^*,l),
\label{OmegaAppendix}
\end{eqnarray}
where
\begin{eqnarray}
A^*_{1+2}(3^*,l_2,l)&=&\frac{1}{\kappa_3^*}\frac{
	\delta^4(k_3+\vll_{l_2}\vlt_{l_2}+\vll_{l}\vlt_{l})
	\delta^8(\vll_{p_3}\vlet_{p_3}+\vll_{l_2}\vlet_{l_2}+\vll_{l}\vlet_{l})
}{\abr{p_3l_2}\abr{l_2l}\abr{lp_3}}\nonumber\\
A^*_{2+1}(2^*,l,l_1)&=&\frac{1}{\kappa_2^*}\frac{
	\delta^4(k_2+\vll_{l_1}\vlt_{l_1}+\vll_{l}\vlt_{l})
	\delta^8(\vll_{p_2}\vlet_{p_2}+\vll_{l_1}\vlet_{l_1}+\vll_{l}\vlet_{l})
}{\abr{p_2l}\abr{ll_1}\abr{l_1p_2}},
\end{eqnarray}
and $A_{3}^{\overline{MHV}}(l_2,1,l_1)$ is defined in a usual way. Note, that for this particular amplitude we have the following kinematical constraints $\lambda_{l_1}\sim\lambda_{\l_2}\sim\lambda_1$ coming from $\overline{\text{MHV}}$ vertex. Using our standard $k_T$ decomposition of off-shell momenta in terms of a pair of on-shell one we have
\begin{eqnarray}
k_2&=&k_2'+k_2'',~
k_2'=|p_2\rangle \frac{k_2|\xi_2\rangle}{\langle p_2 \xi_2\rangle},~
k_2''=-|\xi_2\rangle \frac{k_2|p_2\rangle}{\langle p_2 \xi_2\rangle},
\nonumber\\
k_3&=&k_3'+k_3'',~
k_3'=|p_3\rangle \frac{k_3|\xi_3\rangle}{\langle p_3 \xi_3\rangle},~
k_3''=-|\xi_3\rangle \frac{k_3|p_3\rangle}{\langle p_3 \xi_3\rangle},
\end{eqnarray}
and off-shell vertexes could be written in terms of the
products of $\mbox{MHV}_4$ on-shell amplitudes and inverse soft factors as:
\begin{eqnarray}
A^*_{1+2}(3^*,l_2,l)&=&
S^{-1}(k'_3,k''_3,l_2)
A^{MHV}_4(k'_3,k''_3,l_2,l)\Big{|}_{\tilde{\eta}_{k_3''}=0},~
S^{-1}(k'_3,k''_3,l_2)=\frac{\langle p_3 \xi_3 \rangle \langle \xi_3 l_2 \rangle}
{\kappa_3^* \langle p_3l_3 \rangle},
\nonumber\\
A^*_{1+2}(2^*,l,l_1)&=&S^{-1}(k'_2,k''_2,l_1)
A^{MHV}_4(k''_2,k'_2,l,l_1)\Big{|}_{\tilde{\eta}_{k_2''}=0},~
S^{-1}(k'_2,k''_2,l_1)=\frac{\langle p_2 \xi_2 \rangle \langle \xi_2 l_1 \rangle}
{\kappa_3^* \langle p_2l_1 \rangle}.\nonumber\\
\end{eqnarray}

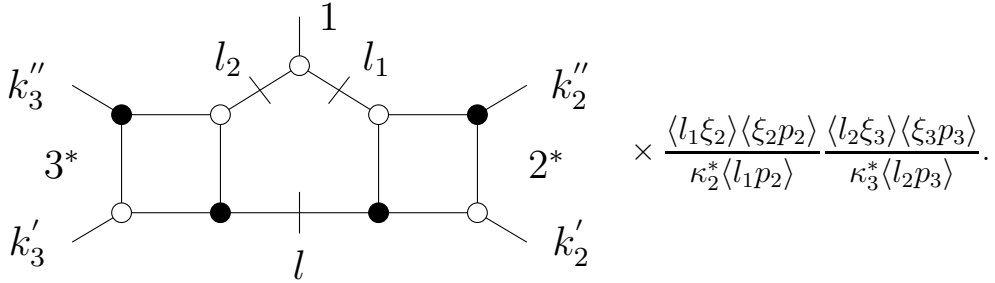
\begin{figure}[htbp]
	\begin{gather*}
	\begin{aligned}
	\begin{tikzpicture}[baseline={($(na.base) - (0,2.5)$)},transform shape,scale=1.3]
	\draw (0,0) -- (1,0);
	\draw (2.6,0) -- (3.6,0);
	\draw (0,-1) -- (3.6,-1);
	\draw (0,0) -- (0,-1);
	\draw (1,0) -- (1,-1);
	\draw (2.6,0) -- (2.6,-1);
	\draw (3.6,0) -- (3.6,-1);
	\draw (1.8,0.5) -- (1,0);
	\draw (1.8,0.5) -- (2.6,0);
	\draw (1.8,0.5) -- (1.8,1);
	\draw (0,0) -- (-0.5,0.3);
	\draw (0,-1) -- (-0.5,-1.3);
	\draw (3.6,0) -- (4.1,0.3);
	\draw (3.6,-1) -- (4.1,-1.3);
	\draw (1.4+0.5*\labelddist,0.25-0.65*\labelddist) -- (1.4-0.5*\labelddist,0.25+0.65*\labelddist);
	\draw (2.2-0.5*\labelddist,0.25-0.65*\labelddist) -- (2.2+0.5*\labelddist,0.25+0.65*\labelddist);
	\draw (1.8,-1-\labelddist) -- (1.8,-1+\labelddist);
	\node[db] (na) at (0,0) {};
	\node[dw] at (1,0) {};
	\node[dw] at (2.6,0) {};
	\node[db] at (3.6,0) {};
	\node[dw] at (0,-1) {};
	\node[db] at (1,-1) {};
	\node[db] at (2.6,-1) {};
	\node[dw] at (3.6,-1) {};
	\node[dw] at (1.8,0.5) {};	
	\node at (1.8+\labelhdist,1) {$1$};	
	\node at (4 + \labelhdist,-0.5) {$2^*$};	
	\node at (-0.3 - \labelhdist,-0.5) {$3^*$};
	\node at (1.4-1.5*\labelddist,0.25+1.5*\labelddist) {$l_2$};
	\node at (2.2+1.8*\labelddist,0.25+1.5*\labelddist) {$l_1$};
	\node at (1.8,-1-2.5*\labelddist) {$l$};
	\node at (4.1+1.5*\labelhdist,0.3) {$k_2^{''}$};
	\node at (4.1+1.5*\labelhdist,-1.3) {$k_2^{'}$};
	\node at (-0.5-1.5*\labelhdist,0.3) {$k_3^{''}$};
	\node at (-0.5-1.5*\labelhdist,-1.3) {$k_3^{'}$};
	\end{tikzpicture}
	\end{aligned}\;\;
	\times\frac{\la l_1\xi_2\ra\la\xi_2 p_2\ra}{\kappa^*_2\la l_1 p_2\ra}\frac{\la l_2\xi_3\ra\la\xi_3 p_3\ra}{\kappa^*_3\la l_2 p_3\ra} .
	\end{gather*}
	\caption{Regulated on-shell diagram for $A^*_{1+2}(1^+,2^*,3^*)$ amplitude
	}
	\label{fig: A1+2-appendix2}
\end{figure}

We may say, that we are "blowing up" off-shell vertexes and express them in terms of the products of four on-shell vertexes and inverse soft factor. This way our on-shell diagram $\Omega$ transforms as depicted in Fig.\ref{fig: A1+2-appendix2}. Now we can use representation of $\mbox{MHV}_3$ and $\overline{\mbox{MHV}}_3$ vertexes as integrals over "small Grassmannians" $Gr(2,3)$ and $Gr(1,3)$ \cite{Henrietta_Amplitudes,AmplitudesPositiveGrassmannian} and integrate out internal on-shell variables. The result of this procedure can be written as \cite{AmplitudesPositiveGrassmannian,Henrietta_Amplitudes}
\begin{eqnarray}
\Omega=\oint \prod_{i=1}^{7}\frac{df_i}{f_i}\;
\delta^{2\times 2} (C (f_i)\cdot\vlt)\delta^{2\times 4} (C (f_i)\cdot\vlet)\delta^{3\times 2} (C^{\perp} (f_i)\cdot\vll)\frac{\la l_1(f_i)\xi_2\ra\la\xi_2 p_2\ra}{\kappa^*_2\la l_1(f_i) p_2\ra}\frac{\la l_2(f_i)\xi_3\ra\la\xi_3 p_3\ra}{\kappa^*_3\la l_2(f_i) p_3\ra},\nonumber\\
\end{eqnarray}
where $\lambda$, $\tilde{\lambda}$ and $\tilde{\eta}$ except $\lambda_{l_1}(f_i)$ and $\lambda_{l_2}(f_i)$ variables may be reconstructed from the ordered set of external on-shell momenta $(k_3',k_3'',1,k_2'',k_2')$ and one
have to put $\tilde{\eta}_{k_2''}=\tilde{\eta}_{k_3''}=0$. The Grassmannian  coordinates $C (f_i)$ could be read of from the on-shell diagram in Fig. \ref{fig: A1+2-appendix2}. The $\lambda_{l_1}(f_i)$ and $\lambda_{l_2}(f_i)$ are some unknown functions of Grassmannian coordinates $f_i$ and external kinematical variables. However, accounting for constraints associated with $\overline{\mbox{MHV}}_3$ vertex we have
\begin{eqnarray}
\frac{\la l_1(f_i)\xi_2\ra}{\la l_1(f_i) p_2\ra}=\frac{\la p_1\xi_2\ra}{\la p_1 p_2\ra},~
\frac{\la l_2(f_i)\xi_3\ra}{\la l_2(f_i) p_3\ra}=\frac{\la p_1\xi_3\ra}{\la p_1 p_3\ra},
\end{eqnarray}
and take inverse soft factors out of the integration sign. Then,
the integration over $f_i$  can be rewritten in the following way:
\begin{eqnarray}
\Omega=\frac{\la p_1\xi_2\ra\la\xi_2 p_2\ra}{\kappa^*_2\la p_1 p_2\ra}\frac{\la p_1\xi_3\ra\la\xi_3 p_3\ra}{\kappa^*_3\la p_1 p_3\ra}~ L_{5}^2 ~\oint\frac{df}{f},\nonumber\\
\end{eqnarray}
where the external kinematical variables in $L_{5}^2$ are again constructed from the ordered set of on-shell momenta $(k_3',k_3'',1,k_2'',k_2')$. Understanding the integration
over $\int df$ as taking residue (i.e. dropping this factorized integral which corresponds to the "bubble reduction" \cite{Henrietta_Amplitudes,AmplitudesPositiveGrassmannian}, see Fig.\ref{fig: A1+2})  we finally obtain (the action of corresponding projectors on $\Omega$ is assumed):
\begin{eqnarray}
\Omega&=&\frac{\la p_1\xi_2\ra\la\xi_2 p_2\ra}{\kappa^*_2\la p_1 p_2\ra}\frac{\la p_1\xi_3\ra\la\xi_3 p_3\ra}{\kappa^*_3\la p_1 p_3\ra} L_{5}^2 =
\frac{\la p_1\xi_2\ra\la\xi_2 p_2\ra}{\kappa^*_2\la p_1 p_2\ra}\frac{\la p_1\xi_3\ra\la\xi_3 p_3\ra}{\kappa^*_3\la p_1 p_3\ra} A^{MHV}_5(k'_2,k_3'',1,k_2'',k_2')\Big{|}_{\tilde{\eta}_{k_2''}=\tilde{\eta}_{k_3''}=0}
\nonumber\\
&=&A^*_{1+2}(1^+,2^*,3^*).
\end{eqnarray}

\end{document}